\pdfoutput=1

\documentclass[11pt,twoside,a4paper,cmspaper,final,collab]{cms-tdr}
\def\svnVersion{461111P}\def\cmsCernNoTag{CERN-EP-2018-015}\def\cmsCernDate{\today}\def\cmsMessage{Published in the Journal of High Energy Physics as \href{http://dx.doi.org/10.1007/JHEP05(2018)088}{\doi{10.1007/JHEP05(2018)088.}}}

\begin{document}\cmsNoteHeader{B2G-16-029}

\hyphenation{had-ron-i-za-tion}
\hyphenation{cal-or-i-me-ter}
\hyphenation{de-vices}
\RCS$Revision: 454693 $
\RCS$HeadURL: svn+ssh://svn.cern.ch/reps/tdr2/papers/B2G-16-029/trunk/B2G-16-029.tex $
\RCS$Id: B2G-16-029.tex 454693 2018-04-09 17:50:17Z clange $

\newcommand{\PV}{\ensuremath{\mathrm{V}}}
\newcommand{\PVt}{\ensuremath{\mathrm{V\!/t}}}
\newcommand{\mWV}{\ensuremath{m_{\PW\xspace \mathrm{V}}}\xspace}%
\newcommand{\mJ}{\ensuremath{m_{\text{jet}}}}%
\newcommand{\nsubj}{\ensuremath{\tau_{21}}}%
\newcommand{\BulkG}{{\ensuremath{\PXXG_{\text{bulk}}}\xspace}\mathchardef\mhyphen="2D}
\renewcommand{\labelenumi}{\roman{enumi}. }

\cmsNoteHeader{B2G-16-029}

\title{Search for a heavy resonance decaying to a pair of vector bosons in the lepton plus merged jet final state at $\sqrt{s} = 13\TeV$}

\date{\today}

\abstract{A search for a new heavy particle decaying to a pair of vector bosons ($\PW\PW$ or $\PW\PZ$) is presented using data from the CMS detector corresponding to an integrated luminosity of 35.9\fbinv collected in proton-proton collisions at a centre-of-mass energy of 13\TeV in 2016. One of the bosons is required to be a \PW\ boson decaying to $\Pe\Pgn$ or $\Pgm\Pgn$, while the other boson is required to be reconstructed as a single massive jet with substructure compatible with that of a highly-energetic quark pair from a \PW\ or \PZ boson decay. The search is performed in the resonance mass range between 1.0 and 4.4\TeV.  The largest deviation from the background-only hypothesis is observed for a mass near 1.4\TeV and corresponds to a local significance of 2.5 standard deviations. The result is interpreted as an upper bound on the resonance production cross section. Comparing the excluded cross section values and the expectations from theoretical calculations in the bulk graviton and heavy vector triplet models, spin-2 $\PW\PW$ resonances with mass smaller than 1.07\TeV and spin-1 $\PW\PZ$ resonances lighter than 3.05\TeV, respectively, are excluded at 95\% confidence level.}

\hypersetup{
pdfauthor={CMS Collaboration},
pdftitle={Search for a heavy resonance decaying to a pair of vector bosons in the lepton plus merged jet final state at sqrt s = 13 TeV},
pdfsubject={CMS},
pdfkeywords={CMS, physics, bulk graviton, diboson, resonances, HVT}}

\maketitle

\section{Introduction}
\label{sec:Introduction}

The discovery of a Higgs boson at the CERN LHC~\cite{Aad:2012tfa,Chatrchyan:2012xdj,Chatrchyan:2013lba} marked the success of
fifty years of scientific investigation, during which the standard
model (SM)~\cite{Glashow:1961tr,Salam:1964ry,Weinberg:1967tq} was
first proposed and then consolidated with experimental evidence.
However, an outstanding issue is the so-called hierarchy problem,
\ie the large difference between the energy scale at which electroweak
symmetry breaks and the Planck scale at which gravity becomes important.
Following a reasoning based on
naturalness~\cite{Susskind:1978ms,tHooft:1979rat,Barbieri:1987fn}, which has been successful in guiding
physics discoveries in the last century~\cite{Giudice:2008bi}, physics
effects beyond the standard model (BSM) are expected at the
electroweak scale.

Different kinds of BSM mechanisms have been proposed to solve the
hierarchy problem. Several of these models predict the existence of new
heavy particles coupled to the vector bosons $\PV=\PW,\PZ$.
Examples include models based on extra spatial
dimensions~\cite{Randall:1999ee,Randall:1999vf,Agashe:2007zd,Fitzpatrick:2007qr}
or on a composite nature of the Higgs
boson~\cite{Agashe:2004rs,Contino:2003ve,Contino:2006nn,Lane:2016kvg}. From previous
searches at
colliders~\cite{Khachatryan:2014hpa,Khachatryan:2014gha,Khachatryan:2015bma,Khachatryan:2015ywa,Khachatryan:2016yji,Khachatryan:2016cfa,Sirunyan:2016cao,Khachatryan:2016cfx,Aad:2015ipg,Aad:2015owa,Aad:2015yza,Aad:2015ufa,Aad:2014xka,Aaboud:2016lwx,Aaboud:2016okv}
and indirect bounds from precision measurements, the masses of these
hypothetical particles (spin-2 bulk gravitons \BulkG\ and spin-1 \PWpr\ and \PZpr\ bosons) are expected to be above
$\approx$1\TeV~\cite{Pappadopulo:2014qza}. With such a large mass, the
resonance decay would result in two bosons of high momentum, which
would give rise to distinctive signatures in the LHC detectors.  For
instance, a high-momentum $\PW$ boson decaying to leptons ($\PW \to
\ell \Pgn$ with $\ell = \Pe, \Pgm$) would lead to the
observation of a lepton aligned with the undetected neutrino.
A vector boson decaying to a $\Pq \Paq^{(')}$ pair
would result in a single, massive jet, which could be
identified using techniques that reveal the substructure of the jet.

\begin{figure}[t!hb]
	\centering
	\includegraphics[width=0.45\linewidth]{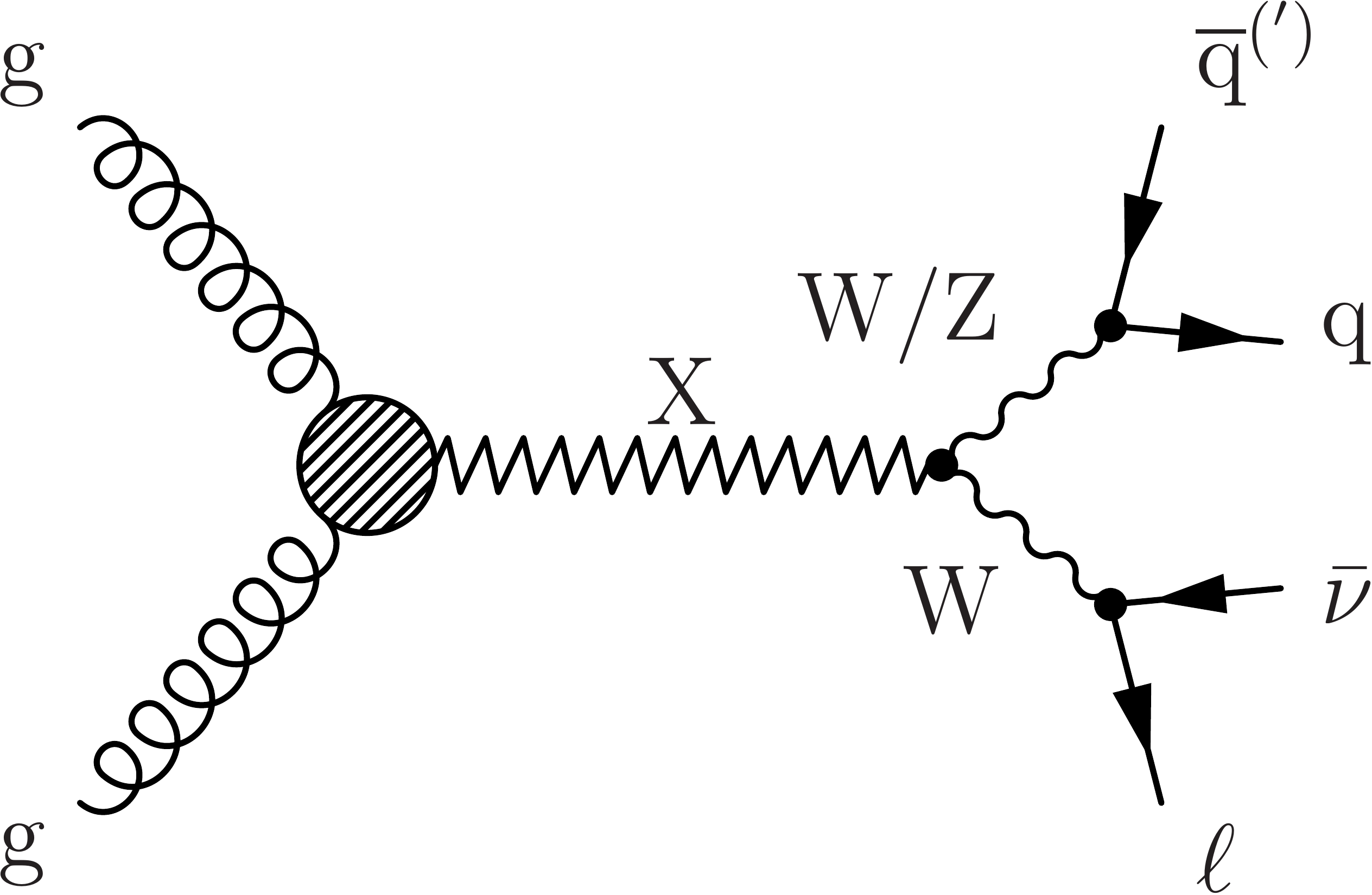}
	\caption{A Feynman diagram for the production of a generic resonance X
		decaying to the $\PW \PW/ \PW \PZ \to \ell \nu \Pq \Paq^{(')}$ final state.}
	\label{fig:FeynDiag}
\end{figure}

In this paper, we describe a search for a heavy resonance decaying to
a pair of vector bosons, one being a $\PW$ boson decaying to an electron or muon
and a neutrino,
the other being a vector boson decaying to a $\Pq \Paq^{(')}$ pair
(see Fig.~\ref{fig:FeynDiag}).  The analysis is based on the
proton-proton collision data set collected by the CMS experiment at the LHC in 2016, at a
centre-of-mass energy of 13\TeV. The collected data
correspond to an integrated luminosity of 35.9\fbinv. Previous
searches for these final states  were performed by the
ATLAS~\cite{Aad:2015ufa,Aaboud:2016okv,Aaboud:2017fgj} and
CMS~\cite{Khachatryan:2014gha,Sirunyan:2016cao} Collaborations,
resulting in stringent bounds on the masses of new resonances, \eg
$m_{\PWpr}>2.99\TeV$ for \PWpr\ particles in the heavy-vector
triplet model B~\cite{Aaboud:2017fgj}, but these searches have no sensitivity for bulk gravitons for the parameters used here (see Section~\ref{sec:models}).

The dominant SM backgrounds to this search arise from lepton+jets events where the jets
either originate from high-momentum vector boson decays to quark-antiquark pairs, $\PV \to \Pq \Paq^{(')}$,
or are ordinary quark- or gluon-initiated jets.
In this analysis, a new signal extraction method using a maximum likelihood fit is deployed,
whereby the SM background contributions are estimated from data during the fit process.
The fit is performed in the plane defined by
the mass of the $\PV \to \Pq \Paq^{(')}$ boson candidate
and the mass of the reconstructed diboson system.
This two-dimensional (2D) approach further exploits the
statistical power of the sideband and signal regions in a simultaneous fit,
improving the discovery sensitivity across a large range of resonance
masses. In addition, the new strategy increases the analysis
flexibility, allowing a simultaneous search for $\PW\PW$ and
$\PW\PZ$ resonances without having to focus on pre-defined jet mass
search windows. The new method is checked against the previously
employed background
prediction method~\cite{Khachatryan:2014gha,Sirunyan:2016cao}, referred to as the
$\alpha$ method.

This paper is organized as follows: the CMS detector is described in
Section~\ref{sec:detector}. Section~\ref{sec:models} introduces the
BSM benchmark models utilized to interpret the
result and the simulated samples used. Section~\ref{sec:selection} describes the event selection.
Section~\ref{sec:backgroundEstimation} discusses the
background model fit and its cross-check, based on the $\alpha$
method. The systematic uncertainties considered are given in Section~\ref{systematics}.
The interpretations of the analysis results in terms of
benchmark BSM models are presented in Section~\ref{sec:results}.
We conclude in Section~\ref{sec:end}.

\section{The CMS detector}
\label{sec:detector}

The central feature of the CMS apparatus is a superconducting solenoid of 6\unit{m} internal diameter, providing a magnetic field of 3.8\unit{T}. Within the solenoid volume are a silicon pixel and strip tracker, a lead tungstate crystal electromagnetic calorimeter (ECAL), and a brass and scintillator hadron calorimeter (HCAL), each composed of a barrel and two endcap sections. Forward calorimeters extend the pseudorapidity ($\eta$) coverage provided by the barrel and endcap detectors. Muons are detected in gas-ionization chambers embedded in the steel flux-return yoke outside the solenoid. A more detailed description of the CMS detector, together with a definition of the coordinate system used and the relevant kinematic variables, can be found in Ref.~\cite{Chatrchyan:2008zzk}.

\section{Benchmark signal models and simulated background samples}
\label{sec:models}
To interpret the results of this study,
two BSM benchmark scenarios are considered: a
spin-2 bulk graviton model in which \BulkG\ decays to $\PW\PW$~\cite{Agashe:2007zd} and a heavy vector triplet (HVT) model with
a charged spin-1 \PWpr\ decaying to $\PW\PZ$~\cite{Pappadopulo:2014qza}.
For the bulk graviton interpretation, the ratio of the unknown curvature scale of the extra dimension, $k$, and the reduced Planck mass, $\overline{M}_{\mathrm{Pl}} \equiv M_{\mathrm{Pl}}/\sqrt{8\pi}$, is set to $\tilde{k} \equiv k/\overline{M}_{\mathrm{Pl}} = 0.5$. This parameter choice ensures that the graviton natural width is negligible with respect to the experimental resolution (narrow-width approximation)~\cite{Oliveira:2014kla}.
The HVT model is a generic framework incorporating several models that predict additional gauge bosons, including composite Higgs models~\cite{Bellazzini:2014yua,Contino:2011np,Marzocca:2012zn,Greco:2014aza,Lane:2016kvg}, which are relevant to this analysis. The specific models are expressed in terms of a few parameters: the strength of the couplings to fermions, $c_\mathrm{F}$, the strength of the couplings to the Higgs boson and longitudinally polarized SM vector bosons, $c_\mathrm{H}$, and the interaction strength $g_\mathrm{V}$ of the new vector boson. For the analysis presented here, samples were simulated in HVT model B, corresponding to $g_\mathrm{V}=3$, $c_\mathrm{H}=-0.98$, and $c_\mathrm{F}=1.02$~\cite{Pappadopulo:2014qza}. For these parameters, the new resonances are narrow and have large branching fractions to boson pairs, while the fermionic couplings are suppressed.
For each hypothesis, we consider resonance masses in the range 1.0--4.4\TeV.  Simulated signal events are generated at leading order (LO)
accuracy with \MGvATNLO v2.2.2~\cite{Alwall:2014hca} with a
relative resonance width of 0.1\%.
The LO production cross section for \BulkG\ resonances is rescaled
by a mass-dependent $K$-factor, to match next-to-leading order (NLO) cross section
values~\cite{Oliveira:2014kla}.

SM background samples are generated using Monte Carlo (MC) simulation.
The \PW+jets process is simulated with
\MGvATNLO at LO and normalized to the next-to-next-to-leading-order (NNLO) cross section, computed using \FEWZ v3.1~\cite{Li:2012wna}. The $\PW$ boson transverse momentum (\pt) spectrum is corrected to account for NLO quantum chromodynamics (QCD) and electroweak contributions~\cite{Kallweit:2015dum}.
Top quark-antiquark (\ttbar{}) events are generated with \POWHEG
v2~\cite{Nason:2004rx,Frixione:2007vw,Alioli:2010xd,Alioli:2009je,Re:2010bp,Alioli:2011as}
and rescaled to the NNLO cross section value computed with \textsc{Top++} v2.0~\cite{Czakon:2011xx}.
Single top quark events are generated with both
\MGvATNLO
($s$-channel) and \POWHEG (associated tW and $t$-channel production) at NLO,
while diboson processes are generated at NLO with
\MGvATNLO
using the merging scheme in Ref.~\cite{Frederix:2012ps}
for \PW\PZ and \PZ{}\PZ, and with \POWHEG for \PW\PW.
The simulated single top quark and diboson background is normalized using inclusive
cross sections calculated at NLO, or
NNLO in QCD, where available, using
\MCFM v6.6~\cite{MCFM:VV,Campbell:2012uf,MCFM:SingleTop}.
Parton showering and hadronization are implemented through
\PYTHIA v8.205 \cite{Sjostrand:2006za,Sjostrand:2007gs} using the
CUETP8M1 tune (CUETP8M2 for \ttbar{}
samples)~\cite{Skands:2014pea,Khachatryan:2015pea}.  The NNPDF
3.0~\cite{Ball:2011mu} parton distribution functions (PDFs) are used
for all simulated samples.
All events are processed through a
\GEANTfour-based~\cite{Agostinelli:2002hh} simulation of the CMS
detector.

Simulated minimum bias interactions are added to the
generated events to match the additional particle production observed
in the large number of overlapping proton-proton interactions within
the same or adjacent bunch crossings (pileup). The simulated events are weighted to reproduce the distribution of the number of pileup interactions observed in data, with an average of 21 reconstructed collisions per beam crossing.
Furthermore, the simulated events are
corrected for differences between data and simulation in the
efficiencies of the triggers, lepton
identification and isolation~\cite{CMS:FirstInclZ}, and selection of
jets originating from hadronization of b quarks (b jets)~\cite{BTV-16-002}.

\section{Event reconstruction and selection}
\label{sec:selection}

Event reconstruction is based on the particle-flow (PF) algorithm~\cite{Sirunyan:2017ulk}, which reconstructs
and identifies each individual particle with an optimized combination of
information from the various elements of the CMS detector.
All events are required to have at least one primary vertex reconstructed within a 24\cm window along the beam axis, with a transverse distance from the nominal pp interaction region of less than 
2\cm~\cite{Chatrchyan:2014fea}. The reconstructed vertex with the largest value of summed physics-object $\pt^2$ is taken to be the primary $\Pp\Pp$ interaction vertex. The physics objects are the jets, clustered using the jet finding algorithm~\cite{Cacciari:2008gp,Cacciari:2011ma} with the tracks assigned to the vertex as inputs, the charged leptons, and the associated missing transverse momentum, taken as the negative vector sum of the \pt of those jets and leptons.

The curvature of muon tracks is obtained by a global fit using measurements from the inner tracker and the muon
detectors. The energy of electrons is determined
from a combination of the electron momentum at the primary interaction
vertex as determined by the tracker, the energy of the corresponding
ECAL cluster, and the energy sum of all bremsstrahlung photons
spatially compatible with originating from the electron track. The energy of charged hadrons is determined from a combination
of their momentum measured in the tracker and the matching ECAL and
HCAL energy deposits, corrected for zero-suppression effects and for
the response function of the calorimeters to hadronic
showers. The energy of neutral hadrons is obtained from the
corresponding corrected ECAL and HCAL energy.
The missing transverse momentum vector \ptvecmiss is defined as the projection onto the plane perpendicular to the beam axis of the negative vector sum of the momenta of all reconstructed PF objects in the event, and its magnitude is denoted as \ptmiss.

Events are selected by the trigger system~\cite{Khachatryan:2016bia} if a muon is
present in the event with $\pt>50\GeV$ and  $\abs{\eta}<2.4$, or if an electron is
identified within $\abs{\eta}<2.5$ with thresholds of $\pt>27$, 55 and
105\GeV for tight, loose, or no isolation criteria applied~\cite{Khachatryan:2015hwa}, respectively.
In addition, events with $\ptmiss>120\GeV$ are
included to further increase the trigger efficiency by exploiting
the high \pt of the neutrino present in the leptonic $\PW$ boson decay.

The offline muon and electron event selection requires $\pt>55\GeV$ with the same $\eta$-acceptance cuts as applied in the trigger.
Requirements on lepton reconstruction quality and lepton identification are
optimized to maintain a high reconstruction efficiency over the whole
energy spectrum~\cite{Chatrchyan:2012xi,Khachatryan:2015hwa}. The muons are required to be isolated from other
particles by requiring that the $\pt$ sum of charged and pileup-corrected neutral particles in a cone of $\Delta R = \sqrt{\smash[b]{(\Delta\eta)^2+(\Delta\phi)^2}} = 0.3 $ (where the azimuthal angle $\phi$ is measured in radians) around the muon direction is less than 5\% of the muon \pt, to reject muons from heavy-flavour processes and decays in flight.
For electrons, the selection cuts include requirements on the geometrical matching between ECAL depositions and the positions
of reconstructed tracks, the ratio of the energies deposited in the HCAL and ECAL, the distribution
of the ECAL depositions, and the number of reconstructed hits in the silicon tracker.
Requirements on the impact parameters of electron and muon tracks with respect to the primary interaction vertex are applied to suppress the contributions from secondary decays and pileup interactions.
Events with only one identified electron or muon are considered; those with additional muons (electrons) with $\pt>20 (35)\GeV$ are discarded.

Two kinds of jets are clustered: large-radius jets are formed by clustering
the PF particles with the anti-\kt
algorithm~\cite{Cacciari:2008gp, Cacciari:2011ma} using a distance
parameter $R=0.8$, while for standard jets $R=0.4$ is used.
Large-radius jets with $\pt>200\GeV$ and standard jets with $\pt>30\GeV$ are
considered.
The jet momentum is determined as the vectorial sum of all particle
momenta in the jet, and is found from simulation to be within 5 to
10\% of the true momentum over the whole \pt spectrum and detector
acceptance.
Jet energy corrections are derived from simulation, and are confirmed
with in situ measurements of the energy balance in dijet, multijet, $\gamma$+jet, and leptonically decaying $\PZ$+jets
events~\cite{Khachatryan:2016kdb}. 
To suppress jets originating from pileup interactions and to mitigate the impact of pileup on jet-related observables, 
we take advantage of the PUPPI algorithm~\cite{Bertolini:2014bba}, which uses local shape information to rescale the momentum of each particle according to its compatibility with the primary interaction vertex.
Quality criteria are applied to the jets
to remove spurious jet-like features originating from
isolated noise~\cite{CMS-PAS-JME-16-003}. In addition, global filters are applied to remove
events with instrumental noise, which would result in artificially large
values of \ptmiss~\cite{CMS-PAS-JME-16-004}. Both standard and large-radius jets are required to lie within the tracker acceptance of $\abs{\eta}<2.5$
where the pileup jet identification and jet substructure algorithms have optimal performance.
Since the signal is expected to be produced centrally, this angular requirement has no significant effect on the signal selection efficiency.
Large-radius jets located within $\Delta R<1.0$ of a selected lepton are discarded, as well as standard jets located within $\Delta R<0.8$ of a large-radius jet or $\Delta R<0.4$ of a selected lepton.

Events of the muon (electron) channel are considered in the analysis
if the \ptmiss in the event is greater than $40 (80)\GeV$.
The \ptvecmiss is considered as an estimate of the \ptvec of the neutrino coming from the $\PW$ boson decay,
and the longitudinal component $p_z$ of the neutrino momentum is estimated
by imposing a $\PW$ boson mass constraint to the lepton+neutrino system
and solving the corresponding quadratic equation.
The solution with smallest magnitude of the neutrino $p_z$ is considered.
When no real solution is found, only the real part is considered.
The leptonically decaying $\PW$ boson candidate is then required to have $\pt>200\GeV$,
and is combined with the most energetic large-radius jet
in the event to form a $\PW\PV$ resonance candidate.

In order to identify large-radius jets as Lorentz-boosted vector bosons, we define a
$\PV$ tagging algorithm, based on an estimate of the jet mass and the
ratio of the $N$-subjettiness~\cite{Thaler:2010tr} variables
$\nsubj = \tau_2/\tau_1$. The jet mass is determined by applying a modified mass-drop algorithm~\cite{Dasgupta:2013ihk,Butterworth:2008iy}, known as the \emph{soft-drop} algorithm~\cite{Larkoski:2014wba}, to large-radius jets, with parameters $\beta=0$,
$z_\text{cut}=0.1$, and $R_0 = 0.8$. The $N$-subjettiness variables are computed fixing
the values of the input parameters to $\beta=1.0$ and $R_0=0.8$.
Jets coming from two-prong W or Z decays are characterized by lower values of $\nsubj$ than one-prong jets from SM backgrounds.
Large-radius jets with soft-drop mass $\mJ$ between
30~and~$210\GeV$ and having $\nsubj <0.75$ are tagged as $\PV$
jets. The use of a large window for $\mJ$ allows $\PW$, $\PZ$, and boosted top quark
large-radius jet candidates to be selected, while retaining a sizeable low-mass sideband for
background characterization.

Standard jets originating from b~quarks are identified by applying the combined secondary vertex algorithm (CSVv2)~\cite{BTV-16-002}. Events are rejected if a selected standard jet passes the medium working point of this algorithm. This working point has a probability for light-flavour jets (attributed to u, d, s, or g partons) to be misidentified as b jets of about 1\%, and a b jet identification efficiency of about 70\%.

The lepton is required to be well separated from the $\PV$-tagged jet, requiring an
angular distance $\Delta R>\pi/2$ between them.
In addition, the difference in azimuthal angle between the $\PV$-tagged jet and both the
\ptvecmiss and the $\PW \to \ell \nu$ boson candidate directions is required to be $\Delta \phi > 2$.
The diboson mass $\mWV$ is computed from the sum of the four-momenta
of the $\PW\to \ell \Pgn$ boson candidate and the \PV-tagged jet. Events
with $\mWV>800\GeV$, for which a monotonically decreasing \mWV spectrum is guaranteed, are considered in the analysis.
The overall selection efficiency times acceptance ranges from 47 to 57\% for the \BulkG\ signal, and from 45 to 60\% for the \PWpr\ signal, increasing with resonance mass.

The selected sample is separated into four mutually exclusive
categories.
First, the sample is split by lepton flavour, distinguishing muon from electron
events. This facilitates accounting for the differences introduced by the
different lepton reconstruction and selection. Subsequently, events are classified as
high-purity (HP) or low-purity (LP), by requiring the $\PV$-tagged jet to
have $\nsubj \leq 0.55$ or $0.55<\nsubj\leq 0.75$,
respectively. The definition of the HP and LP event categories was
optimized by maximizing the expected significance for a bulk graviton signal over the full mass range using simulated events with $65<\mJ<105\GeV$.

\begin{figure}[h]
	\centering
	\includegraphics[width=0.45\textwidth]{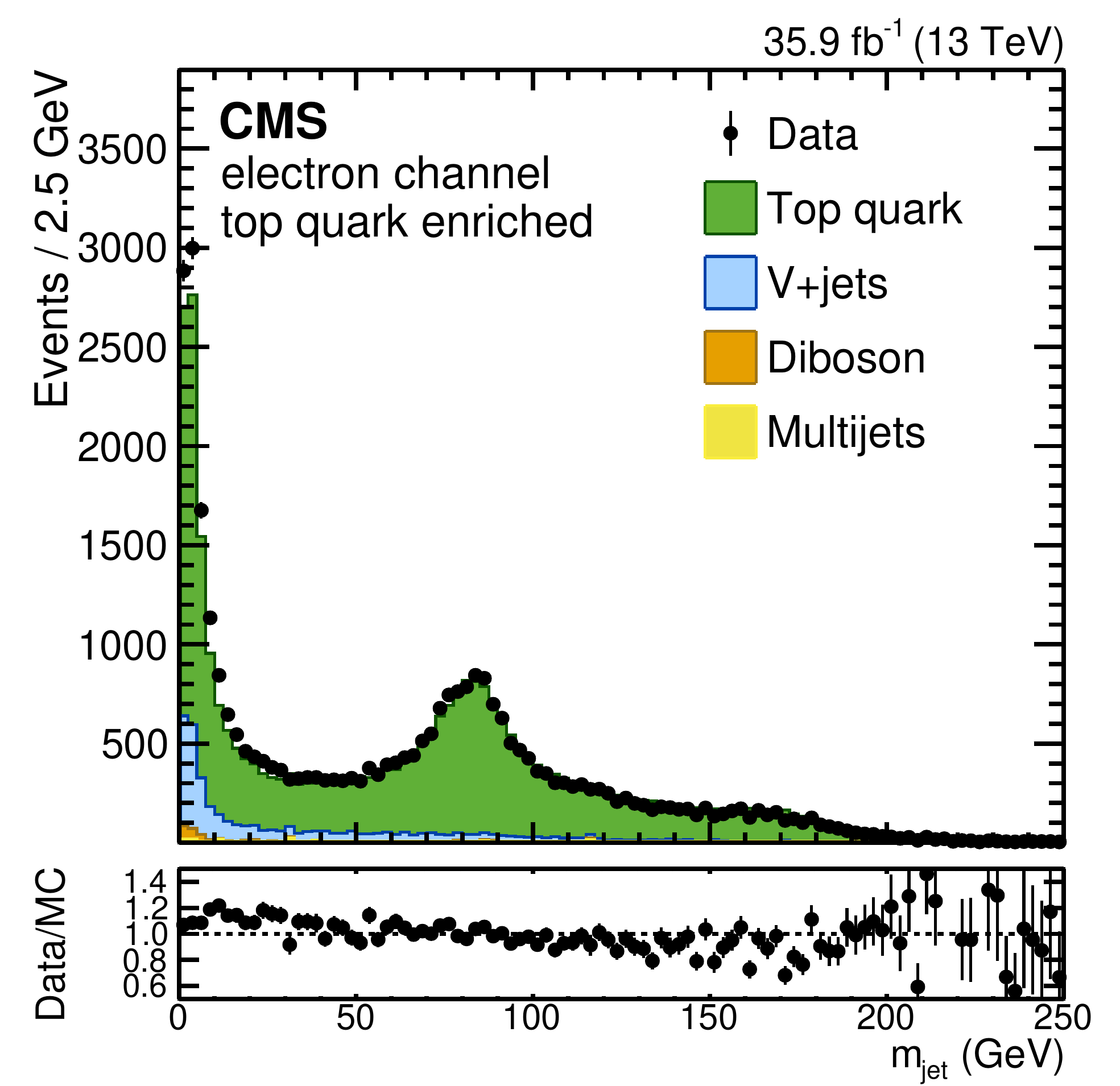}
	\includegraphics[width=0.45\textwidth]{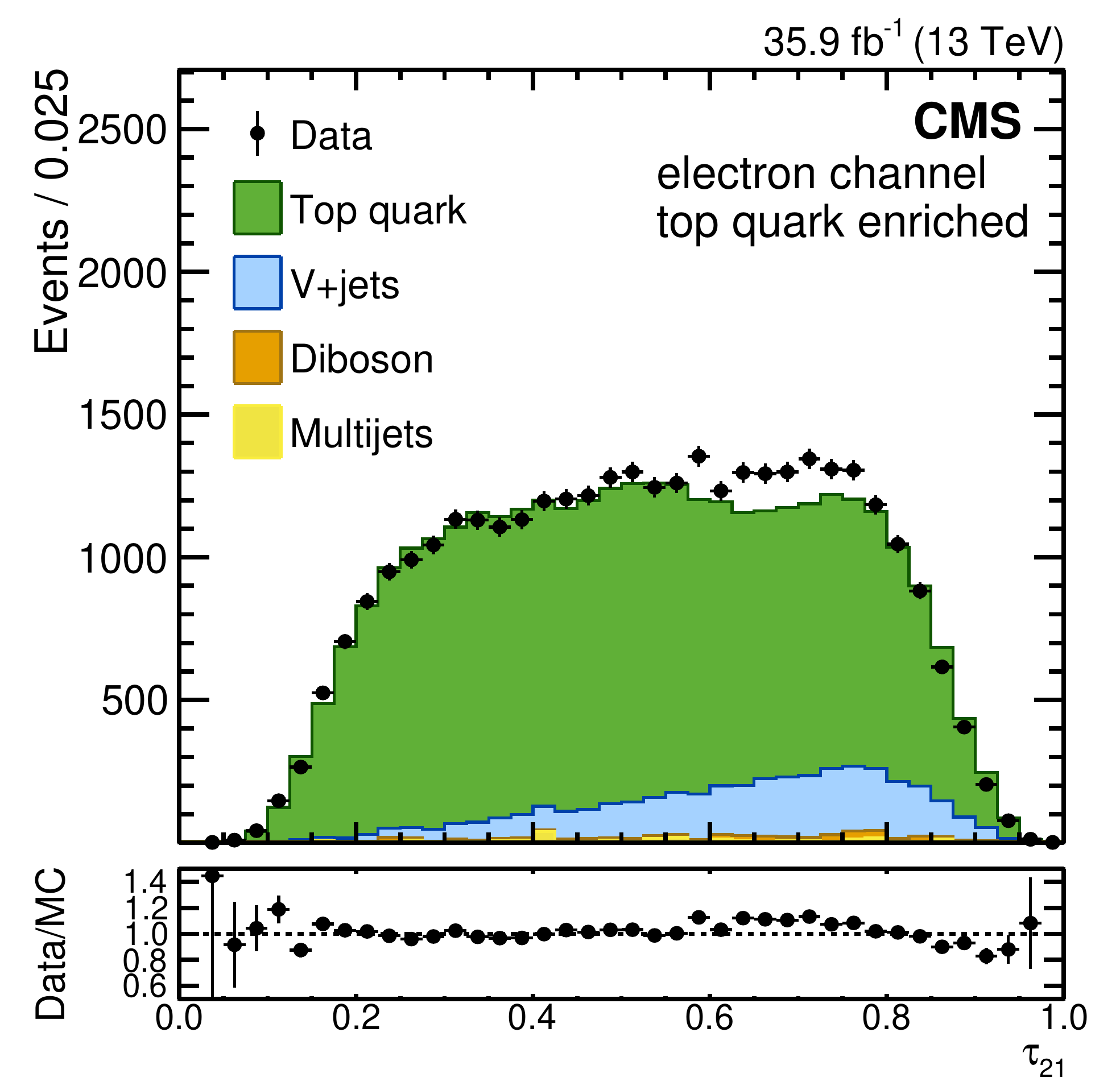}
	\caption{Jet soft-drop mass (left) and $N$-subjettiness ratio
		$\nsubj$ (right) for data and simulated events in the
		top quark enriched region in the electron channel. The contribution labelled as ``Top quark" includes \ttbar{} and single top processes, and the ``V+jets" contribution is dominated by \PW+jets events with a small contribution from Z+jets events. The vertical bars correspond to the statistical uncertainties of the data.\label{fig:topControl}}
\end{figure}

A control sample of \ttbar{} events with similar kinematic distributions to the events in the signal region is selected by inverting the b jet veto. This sample is used to quantify the agreement between data and simulation in describing the $\mJ$ and $\nsubj$ variables. Figure~\ref{fig:topControl} shows the distributions of these variables in the b-enriched sample, as an example in the electron channel. The observed disagreement between data and simulation is reasonable, since both variables are sensitive to hadronization, which is difficult to model. After applying the jet mass window and $\nsubj$ selection cuts, this sample is used to derive scale factors for the efficiency of the $\nsubj$ selection, and the resolution and scale of the jet mass peak. Consequently, a correction factor to the \ttbar{} event rate of $0.88 \pm 0.12$ is used, which is not applied in Fig.~\ref{fig:topControl}.

\section{Signal extraction}
\label{sec:backgroundEstimation}

For this analysis, a novel signal extraction method based on a
2D maximum likelihood fit is introduced. As a cross-check,
the prediction obtained with the $\alpha$ method, used in previous versions of this
analysis~\cite{Khachatryan:2014gha,Sirunyan:2016cao}, is also presented.

The signal and background yields are determined through
a maximum likelihood fit, performed in the
portion of the ($\mWV$, $\mJ$) plane defined by the event
selection described in Section~\ref{sec:selection}.
The fit is performed using 2D templates for signal and
background processes, starting from simulation and introducing shape uncertainties that model the
difference between data and simulation in the full search range.

The probability density function (pdf) of $X \to \PW \PV$ events in
the ($\mWV$, $\mJ$) plane is modelled as:
\begin{equation}
P_\mathrm{sig}(\mWV,\mJ|m_{\mathrm{X}}) = P_{\PW \PV}(\mWV|m_{\mathrm{X}},\theta_1) \, P_{j}(\mJ|m_{\mathrm{X}},\theta_2).
\end{equation}
The pdfs $P_{\PW \PV}$ and $P_j$ are represented by double
Crystal Ball~\cite{Oreglia:1980cs,Gaiser:1982yw} functions, and an additional exponential function is used in the jet mass dimension
in LP events to model the tails of the soft-drop jet mass distribution. The parameters of the functions
are described by uncorrelated polynomial interpolations, obtained by fitting the simulated signal sample distributions with the pdfs for different values of the resonance mass $m_{\mathrm{X}}$. The experimental resolution for $\mJ$ is around 10\%, and for $\mWV$ it ranges from 6\% at 1\TeV to 4\% at 4\TeV.

For the signal, the dependence of the shape parameters on the
resonance mass is found to depend on the nature of the $\PV$ jet (\eg
$\PW$ or $\PZ$) and the lepton flavour. The signal yields in the different
signal categories are expressed as a function of the integrated
luminosity of the sample and the product of the signal acceptance and efficiency, treated as nuisance
parameters, so that the resonance production cross section is
determined in a combined fit to data in the four categories.

Two classes of background events are considered:
\begin{enumerate}
	\item A \PW+jets background, consisting of a lepton and at least one jet arising from a quark or gluon
	mistagged as a $\PV$ jet. In addition to $\PW\to \ell \Pgn$+jets, this background also includes \ttbar{}
	production where the leptonically decaying $\PW$ boson was reconstructed, but the merged jet
	corresponds to a random combination of jets in the event and not to a $\PW$ boson or
	a top quark decay.
	\item A $\PW$+$\PVt$ background, peaking in $\mJ$ while smoothly falling in $\mWV$.
	This background is dominated by \ttbar{} production while sub-dominant
	contributions include SM diboson and single top production.
\end{enumerate}
Each background is modelled by a separate shape pdf based on its properties.

The $\PW$+jets background shape is described as a conditional probability
of $\mWV$ as a function of $\mJ$:
\begin{equation}
P_{\PW+\text{jets}}(\mWV,\mJ) = P_{\PW \PV}(\mWV|\mJ,\theta_1) \, P_{j}(\mJ|\theta_2).
\end{equation}
The conditional probability is essential to take into account the large
correlations between $\mJ$ and $\mWV$. Those correlations arise from the
strong dependence of the jet mass on the jet $\PT$ during the hadronization
process.
The 2D conditional templates, $P_{\PW \PV}$, are constructed from simulated events, starting before the detector simulation stage.
For each event in the background samples, jets are clustered from stable particles
using the same substructure algorithms as during event reconstruction. Consequently, a scale and resolution model is derived for both $\mJ$ and
$\mWV$ as a function of generated jet \pt by comparing the reconstructed and generated variables. Smooth templates are then populated as sums of 2D Gaussian distributions, where
the mean values of the Gaussians correspond to the true value of $\mJ$ and $\mWV$, shifted by the derived scale model, and the 2D covariance matrix
is given by the resolution model.
This technique is similar to the kernel-estimation procedure given in Ref.~\cite{Cranmer:2000du} but uses the simulation and the exact resolution model instead of starting from reconstructed events.
The final step is to smooth the tails for high values of $\mWV$ ensuring there are no empty bins in the templates.
The smoothing is performed by fitting events in each $\mJ$ bin with $\mWV >2.0\TeV$ using an exponential
function and then using the function values to populate the tails for $\mWV>2.5 \TeV$.
The $P_{j}$ shapes are one-dimensional (1D) histograms derived directly from reconstructed simulated events, in contrast to the $P_{\PW \PV}$ shapes discussed above.

For both the $P_{j}$ and $P_{\PW\PV}$ components, nuisance parameters are introduced to account for differences between data and simulation.
The most important difference is attributed to the different \pt spectrum of the jets in the simulation. The template
construction is repeated by adding event weights corresponding to a harder (softer) spectrum, and the pdf is interpolated between these alternative templates. An additional uncertainty lies in the choice of the scale/resolution model, which is estimated by varying the scale as functions of $\mJ$ and $\mWV$.
The derived shapes are found to be in agreement with the simulated events, validating the template construction procedure.
This procedure implicitly assumes that a single component can account for the sum of the \ttbar{} events with an arbitrary fraction of reconstructed $\PW$+$\PV$ jet and \PW+jets
contributions. A variation of the relative fractions is found to
translate into a change in the average \pt spectrum, which is taken into account as a systematic uncertainty.

The $\PW$+$\PVt$ background is modelled as :
\begin{equation}
P_{\PW\mathrm{+}\PVt}(\mWV,\mJ) = P_{\PW \PV}(\mWV|\theta_1) \, P_{j}(\mJ|\mWV,\theta_2).
\end{equation}
In this case, $P_{\PW \PV}$ is a 1D template constructed in the same way as for the $\PW$+jets background, and the smoothing of its tail with an exponential function is performed for $\mWV > 1.2\TeV$.
$P_{j}$ is described by two peaks: a peak around the $\PW$ boson mass dominated by top quark events
where only the $\PW \to \Pq\Paq'$ was reconstructed inside the large-radius jet, and a peak
around the top quark mass where the $\PW$ boson and the b quark decays are merged.
These peaks are modelled by two double Crystal Ball plus one exponential function, whose parameters are described by uncorrelated polynomial functions of \mWV.
The presence of both jet peaks allows additional scrutiny, since the relative fraction
of the two peaks as a function of the resonance mass provides a robust validation of the top quark \pt spectrum convolved
with effects from jet grooming. Different shapes are used in the individual event categories to
account for differences in the event kinematic distributions.
The $\mJ$ distribution for the $\PW$+$\PVt$ background in the region of the $\PW$ boson peak is found to differ
from the corresponding signal distribution. Indeed, this background component
mainly consists of high-momentum \ttbar{} events, in which a part of
the b jet from the $\text{t} \to \PW \text{b}$ decay overlaps with the $\PV$ jet
from the $\PW \to \Pq \Paq'$ decay. This special kinematic configuration induces
a modification in the $\mJ$ shape, which is taken into account using
different functions to describe the $\mJ$
distributions of the signal and the $\PW$+$\PVt$ background.

The background estimation method employed in previous versions of this
analysis is also applied, to cross-check the novel fit method. A full
description of this $\alpha$ method, used to estimate the $\PW$+jets background from data, is presented in
Refs.~\cite{Khachatryan:2014gha,Sirunyan:2016cao}.  An
unbinned fit to the $\mWV$ distribution in data is performed, for
events with $40\leq \mJ <65\GeV$ or $135\leq \mJ <150\GeV$.
In this fit, the other background processes (\ttbar, single top quark, etc.),
which are modelled using functional shapes,
are fixed to the prediction from simulated samples.
Using a transfer factor $\alpha(\mWV)$,
estimated from a ratio derived from simulated samples, the result of the
$\mJ$ sideband fit is extrapolated to the $\PW$ and $\PZ$ signal
regions, defined by requiring $65\leq \mJ<85\GeV$ and $85\leq \mJ <
105\GeV$, respectively. The separate $\PW$ and $\PZ$ boson mass windows double the number of signal categories to eight for the $\alpha$ method.
The 2D fit approach provides an improvement to the expected sensitivity of the search by 20\% compared to the $\alpha$ method, relying exclusively on data to predict the shape and the normalization of the backgrounds.

\section{Systematic uncertainties}
\label{systematics}

Several systematic uncertainties affect the overall normalization and
shape of the signal and backgrounds.  Each effect is modelled by a nuisance parameter, which is profiled in the likelihood minimization.
When specified, the
uncertainty size represents the width of the function used to
constrain the nuisance parameter (a log-normal function for
systematic uncertainties related to normalization, and Gaussian
functions for shape uncertainties).

The signal shape for the 2D fit in the $(\mJ,\mWV)$ plane is
affected by several systematic uncertainties: the jet energy scale (2\% in
the $\mWV$ peak position) and resolution (5\% in the $\mWV$
peak width), \ptmiss energy scale and resolution (2\% in both the $\mWV$ peak
position and width).
Additional nuisance parameters are introduced to
allow for variations of the soft-drop jet mass peak due to the effects of grooming on the scale
and the resolution: a jet mass scale uncertainty of 1\% and a resolution
uncertainty of 2\% are applied. Both values were estimated by fitting the $\PW$ peak
from \ttbar{} events in an orthogonal control region defined by requiring the presence of a b-tagged jet.
The jet mass scale and resolution
are correlated across all resonant components including the signal and the $\PW$+$\PVt$ background.

The signal modelling for both methods is furthermore subject to
uncertainties in the lepton modelling, as the uncertainty in the lepton
energy scale is correlated with the obtained signal efficiency.
Changes in lepton energy are propagated to the reconstructed \ptmiss, and
through the entire analysis.  The relative change in the number of
selected signal events is taken as a systematic uncertainty in the
signal normalization. These uncertainties are
smaller than 1\% for both lepton flavours, and are uncorrelated for different lepton flavours.
In addition, the induced change in
the peak position and width are added as systematic uncertainties in
the distribution of the signal with an effect below 1\%.  The systematic uncertainties in the
lepton trigger, identification, and isolation efficiencies are
obtained using a ``tag-and-probe'' method in $\PZ \to \Pe \Pe / \Pgm \Pgm$
events~\cite{CMS:FirstInclZ}. An uncertainty of 1--3\% is assigned to
the trigger efficiency for both lepton flavours, depending on the
lepton \pt and $\eta$. For identification and isolation
efficiencies, the systematic uncertainty is estimated to be 1--2\% for
muons and 3\% for electrons.

A 2.5\% uncertainty in the integrated
luminosity~\cite{CMS-PAS-LUM-17-001} applies to the normalization of
signal events. Uncertainties in the signal yield due to the choice of
PDFs and factorization and renormalization scales are also taken into account by quantifying the change in acceptance.
The scale uncertainties are evaluated following the proposal in Refs.~\cite{Cacciari:2003fi,Catani:2003zt}. The PDF uncertainties are evaluated using the NNPDF 3.0~\cite{Ball:2011mu} PDF set.
The resulting uncertainties in acceptance are found to be negligible for the scale variation and range from 0.1 to 2\% for the PDF evaluation.
The signal cross section uncertainty arising from the uncertainty in PDFs and scales is not taken into account in the statistical analysis, but instead considered as an uncertainty in the theoretical cross section. These cross section uncertainties vary from 4 to 77\% and from 2 to 23\%, respectively, depending on the resonance mass, particle type, and its production mechanism.

The background normalization and shape are estimated in the fit process. Therefore, large a priori uncertainties are assigned to the corresponding nuisance parameters, which are then constrained by the data.
For the $\PW$+jets background, a normalization uncertainty of 50\% is assumed, even though the observed difference between data and simulation when normalizing to integrated luminosity is significantly smaller.
This large a priori uncertainty has no impact on the sensitivity; it provides a loose initial constraint for the fit, which precisely derives the normalization.
For the $\PW$+$\PVt$ background, which is dominated by \ttbar{} production, a normalization uncertainty of
20\% is assigned. Data and simulation in the top quark enriched region defined by inverting the b jet
veto agree to better than 10\%.  The background normalization uncertainties are not correlated
across the different categories.

Further systematic uncertainties lie in the $\PW$+jets and $\PW$+$\PVt$  $m_{\PW \PV}$ background shapes derived by the template-building method described in Section~\ref{sec:backgroundEstimation}.
These uncertainties are encoded in alternative shape functions, derived by repeating the template construction for different assumptions on the jet \pt spectrum as well as on the jet and resonance mass scale and resolution.
One nuisance parameter is used to account for potential shifts of the $\mWV$ spectrum due to uncertainties in the jet \pt spectrum: each bin along the \mWV direction is shifted by $\pm 0.1 \, \mWV /\TeV$  to create mirrored templates in both directions.
Similarly, another parameter is motivated by the measured dependence of the $\mWV$ scale as a function of jet mass, resulting in a bin-by-bin shift of $\pm 400\GeV/\mWV$.
The soft-drop jet mass shape can be affected by additional physics effects specific to hadronization and jet substructure: two additional nuisance parameters are therefore used, one changing the shape by $\pm 0.002 \, \mJ /\GeV$ and another one by $\pm 15\GeV/\mJ$.
The values of these coefficients are chosen such that the resulting alternative shapes cover any differences between data and simulation observed in control regions.
Several tests were performed with these variations and it was found that adding further parameters does not introduce any significant bias in the signal yields.

The $\PW$+$\PVt$ soft-drop jet mass shape is varied by scale and resolution uncertainties as is done for the signal. Additionally, a variation of the relevant
fraction of the resonant $\PW$ boson and top quark mass peaks is taken into account. By fitting the simulated distributions, we observed that the fraction of the two peaks can be modelled as:
\begin{equation}
f = a + b/{\mWV^2}.
\end{equation}
Two nuisance parameters are introduced to model the change of the top quark \pt spectrum, allowing a variation of $a$ by 0.2 and
of $b$ by $25000\GeV^2$. With these two parameters, the \pt spectrum of the top quark can be varied by about 30\%, constraining the $\PW$ boson and top quark jet mass peaks.

The event categorization based on jet substructure introduces two additional sources of systematic
uncertainties, namely the efficiency associated with the $\nsubj$ requirement (HP: 14\%, LP: 33\%) and the
dependence of this efficiency on the jet \pt (4--13\%). Both effects introduce a migration of events between
the LP and HP categories.
In total, 55 independent nuisance parameters are considered in the 2D fit, 14 of which affect the normalization of the signal and backgrounds, while 41 affect their shapes.

When applying the $\alpha$ method, all the normalization uncertainties
listed above are taken into account. The uncertainty in the $\mWV$
distribution is dominated by the statistical uncertainties in the
simultaneous fits to the data of the sideband region, and the
simulation in sideband and signal regions. An effect of almost equal
magnitude arises from the uncertainties in the modelling of the transfer
function $\alpha(\mWV)$ between the sideband and the signal
region. The uncertainty in the normalization of the background has
three sources: the $\PW$+jets component, dominated by the statistical
uncertainty of the events in the $\mJ$ sideband, varying from 2 to
6\%; the \ttbar/single top quark component, dominated by the scale factor
obtained from the top quark enriched control region, amounting to
about 1--3\%, and by the b tagging scale factor, contributing 2--3\%;
and the diboson component, dominated by the $\PV$ tagging uncertainty,
which varies in the range of 15--35\%.

\section{Results}
\label{sec:results}

\begin{figure}[htbp]
	\centering
		\includegraphics[width=0.45\textwidth]{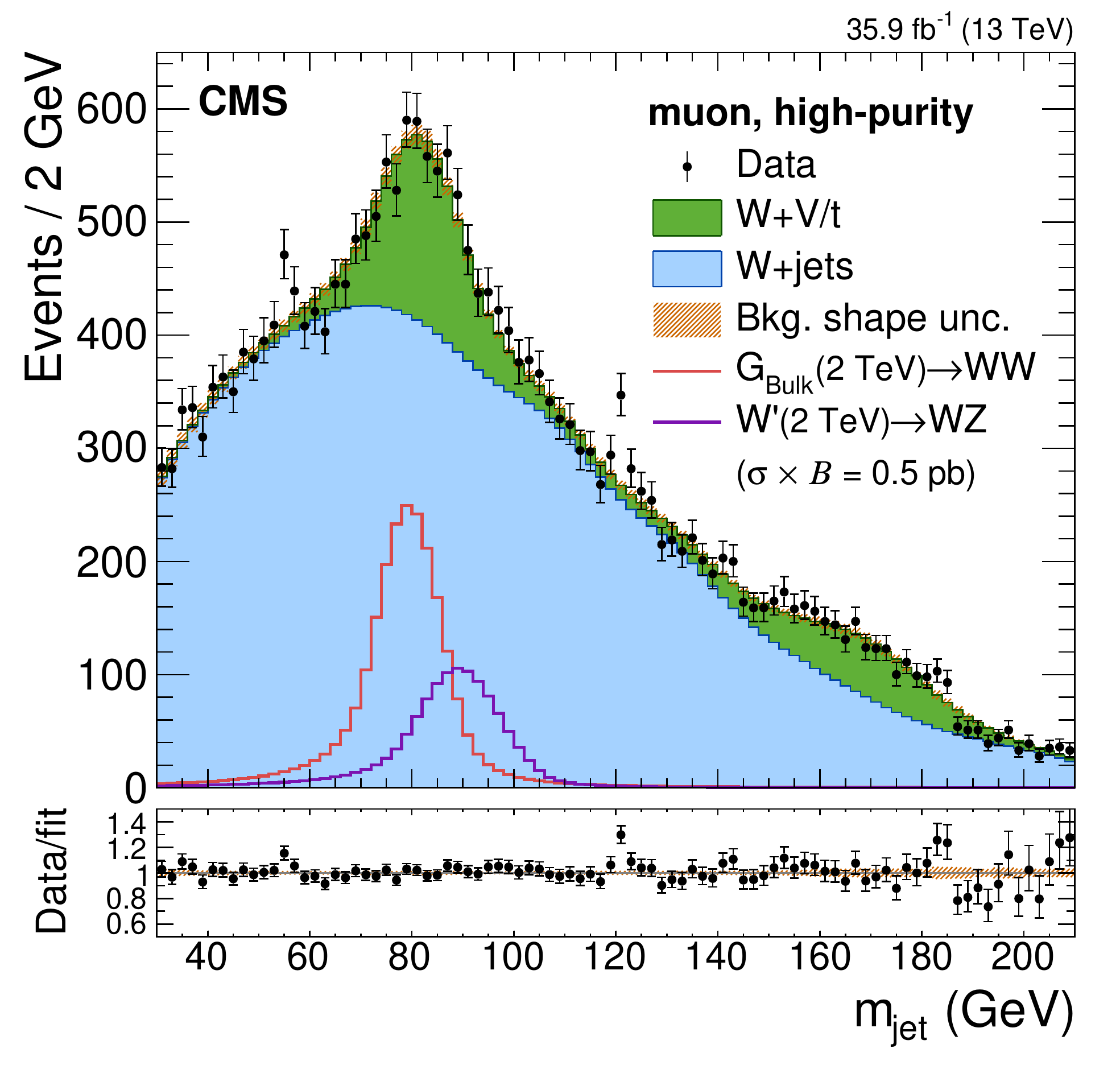}
		\includegraphics[width=0.45\textwidth]{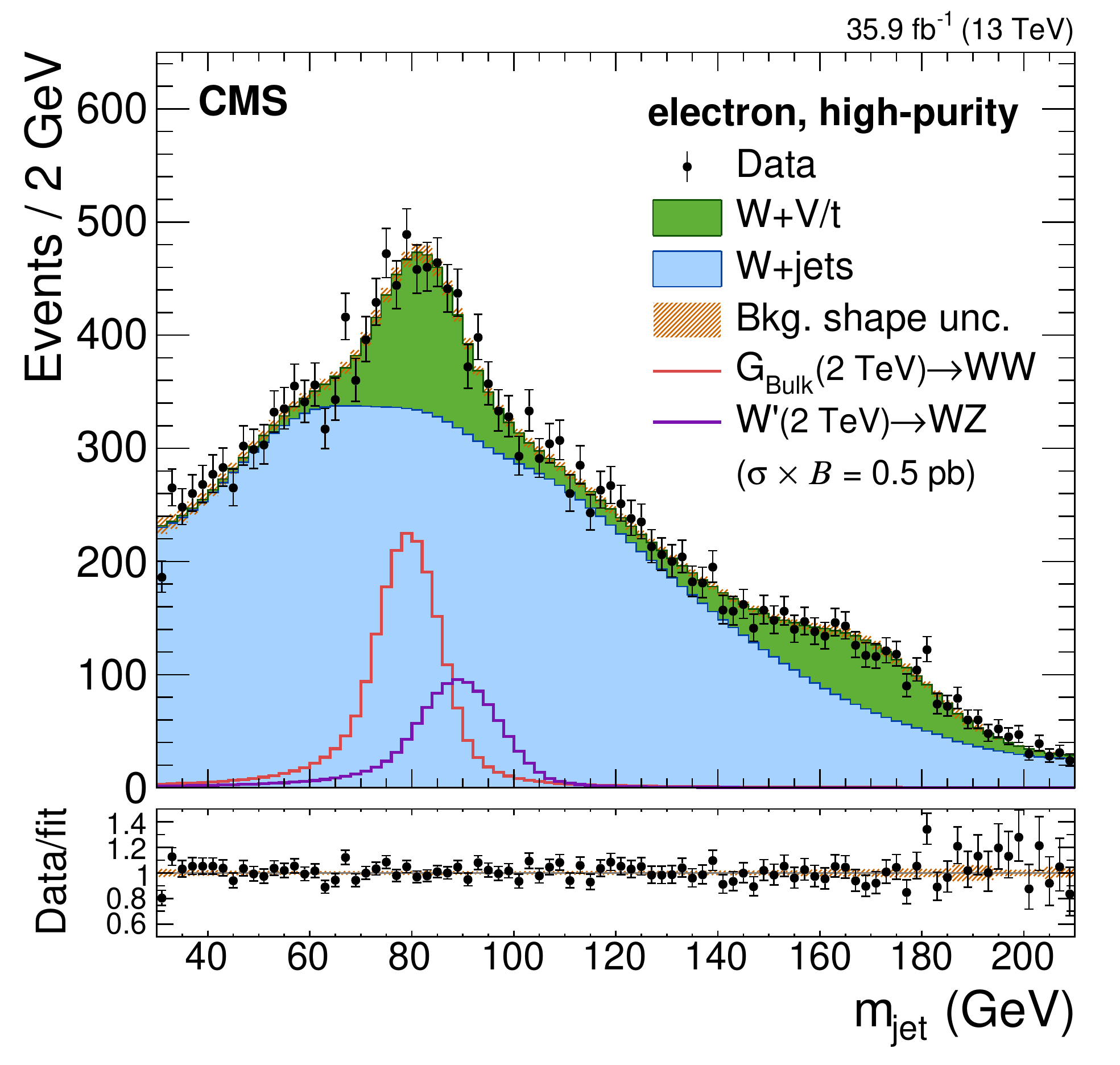}
		\includegraphics[width=0.45\textwidth]{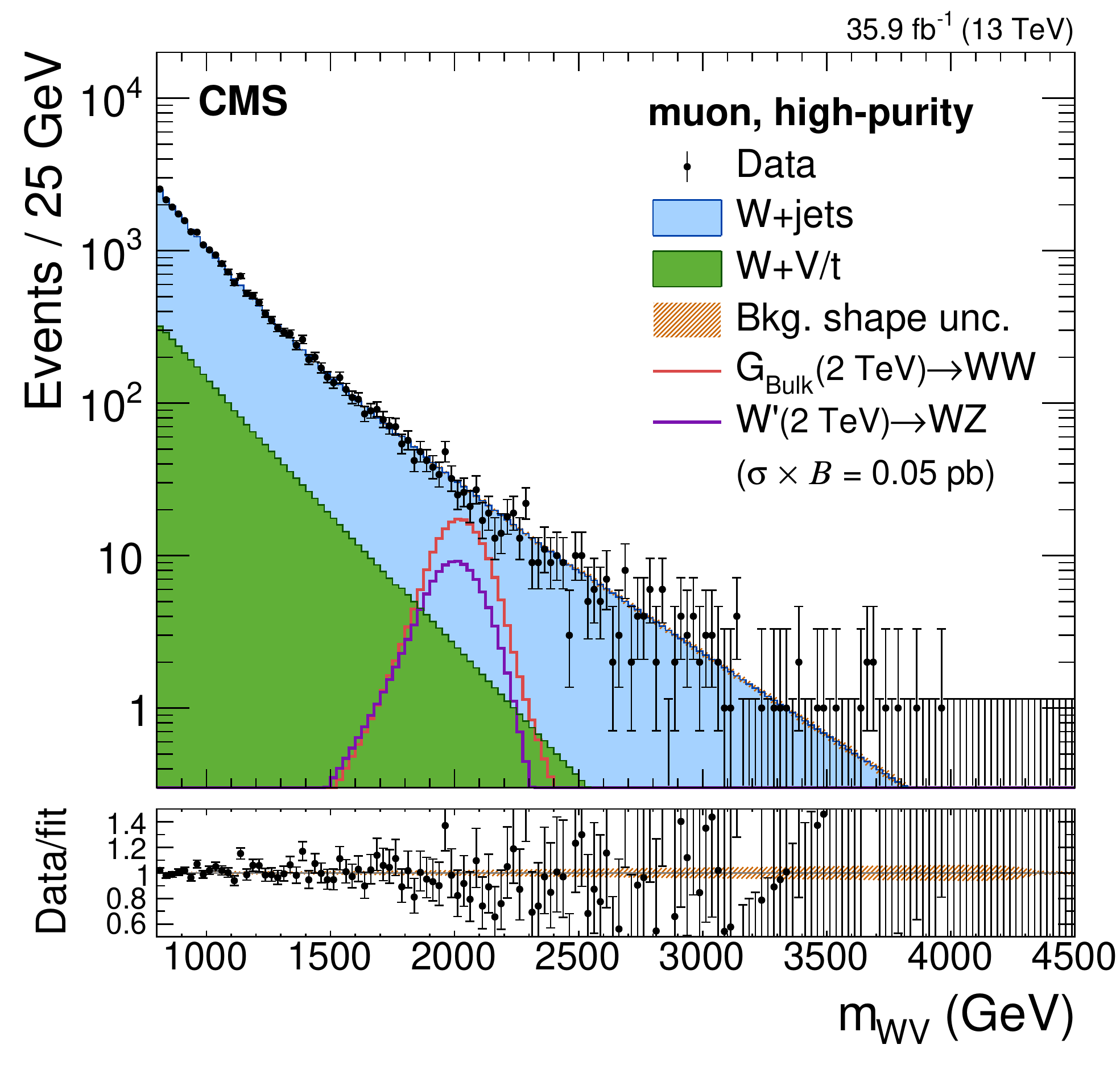}
		\includegraphics[width=0.45\textwidth]{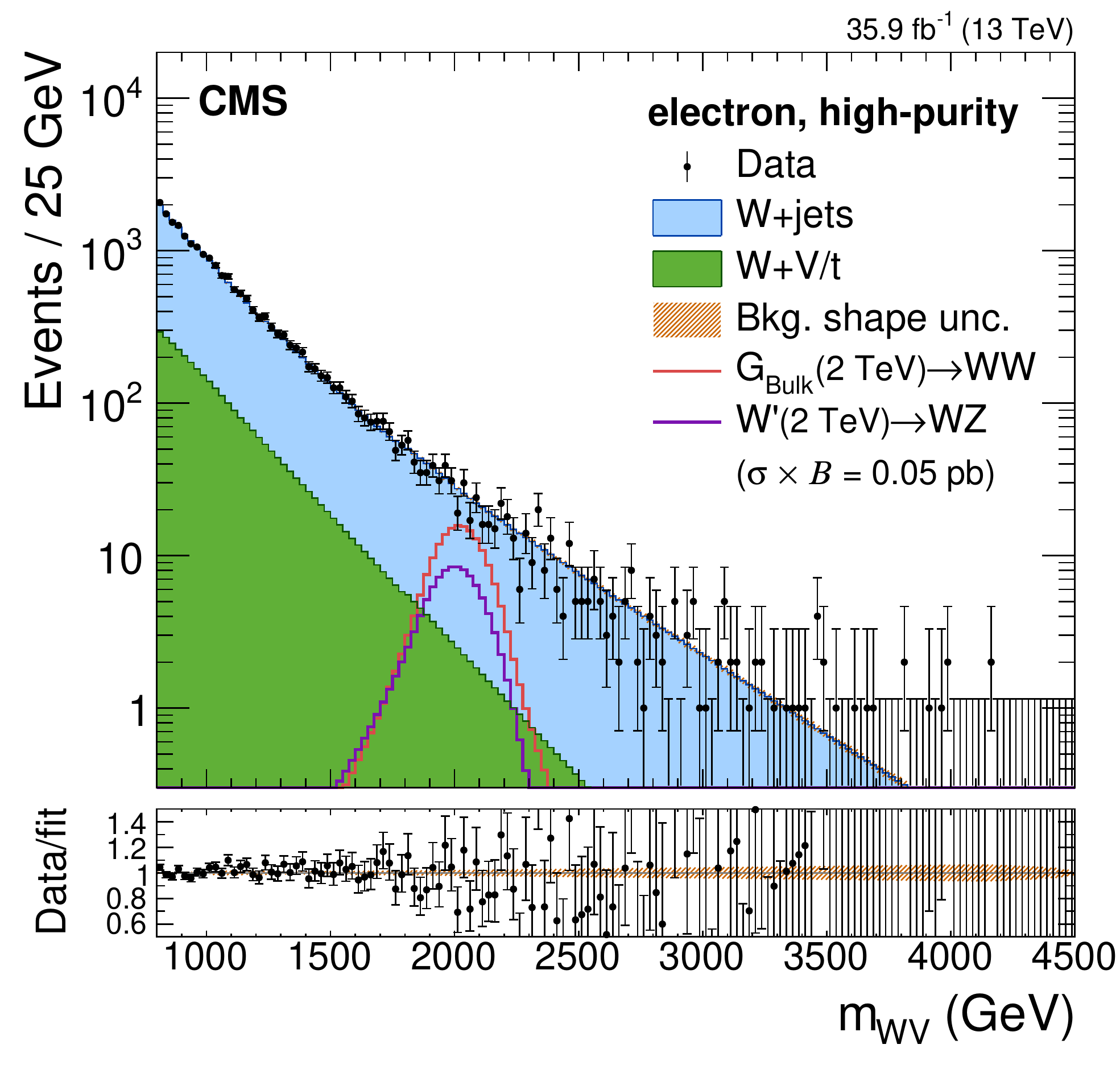}
	\caption{Comparison between the fit result and data distributions of $\mJ$ (upper) and $\mWV$ (lower)
		in the muon HP (left) and electron HP (right) category.
		The background shape uncertainty is shown as a shaded band, and the statistical uncertainties of the data are shown as vertical bars.
		No events are observed with $\mWV>4.5\TeV$.
		Example signal distributions are overlaid, using an arbitrary
		normalization that is different in the upper and lower plots.
		\label{fig:HP}}
	
\end{figure}

\begin{figure}[htbp]
	\centering
		\includegraphics[width=0.45\textwidth]{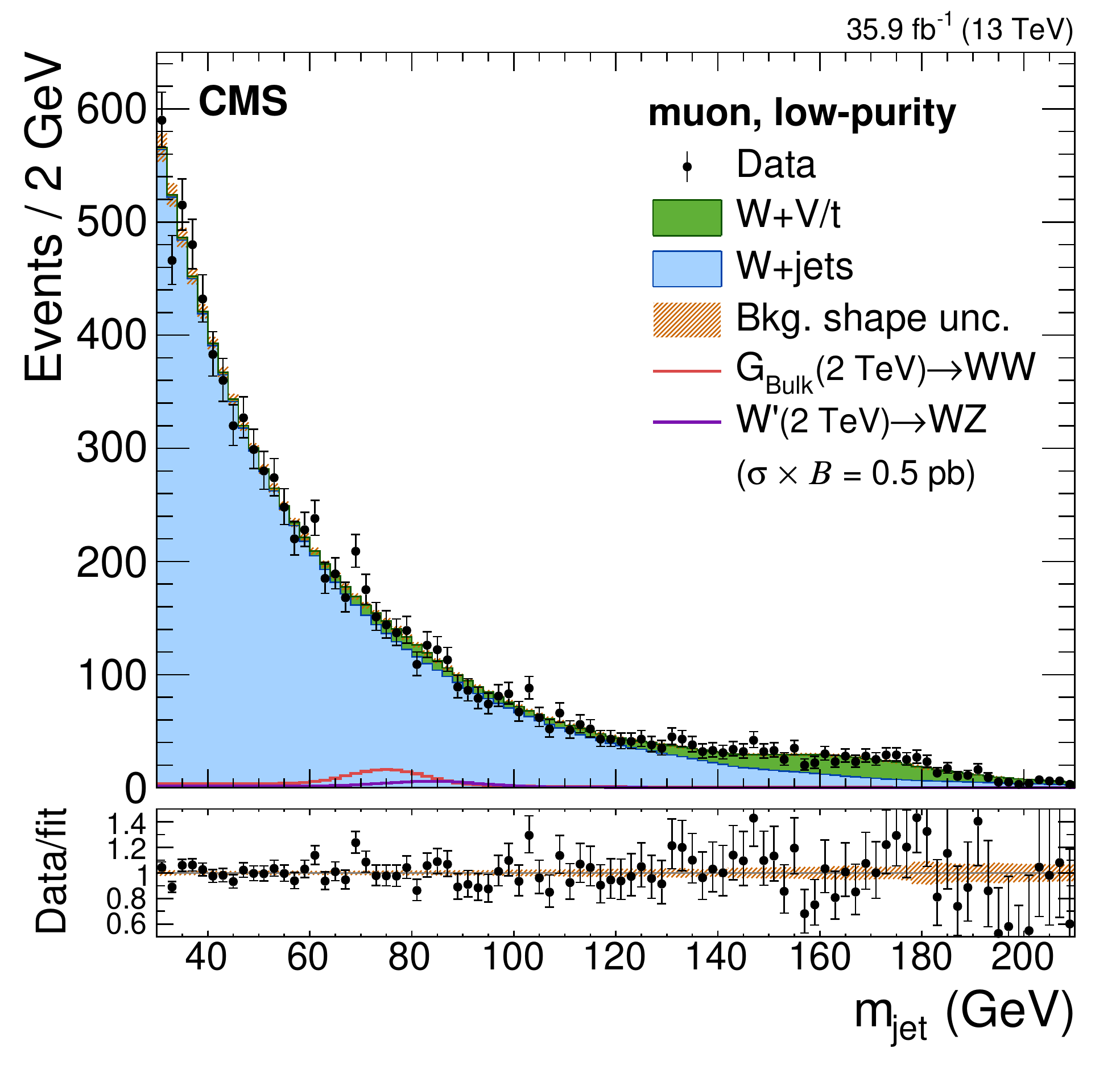}
		\includegraphics[width=0.45\textwidth]{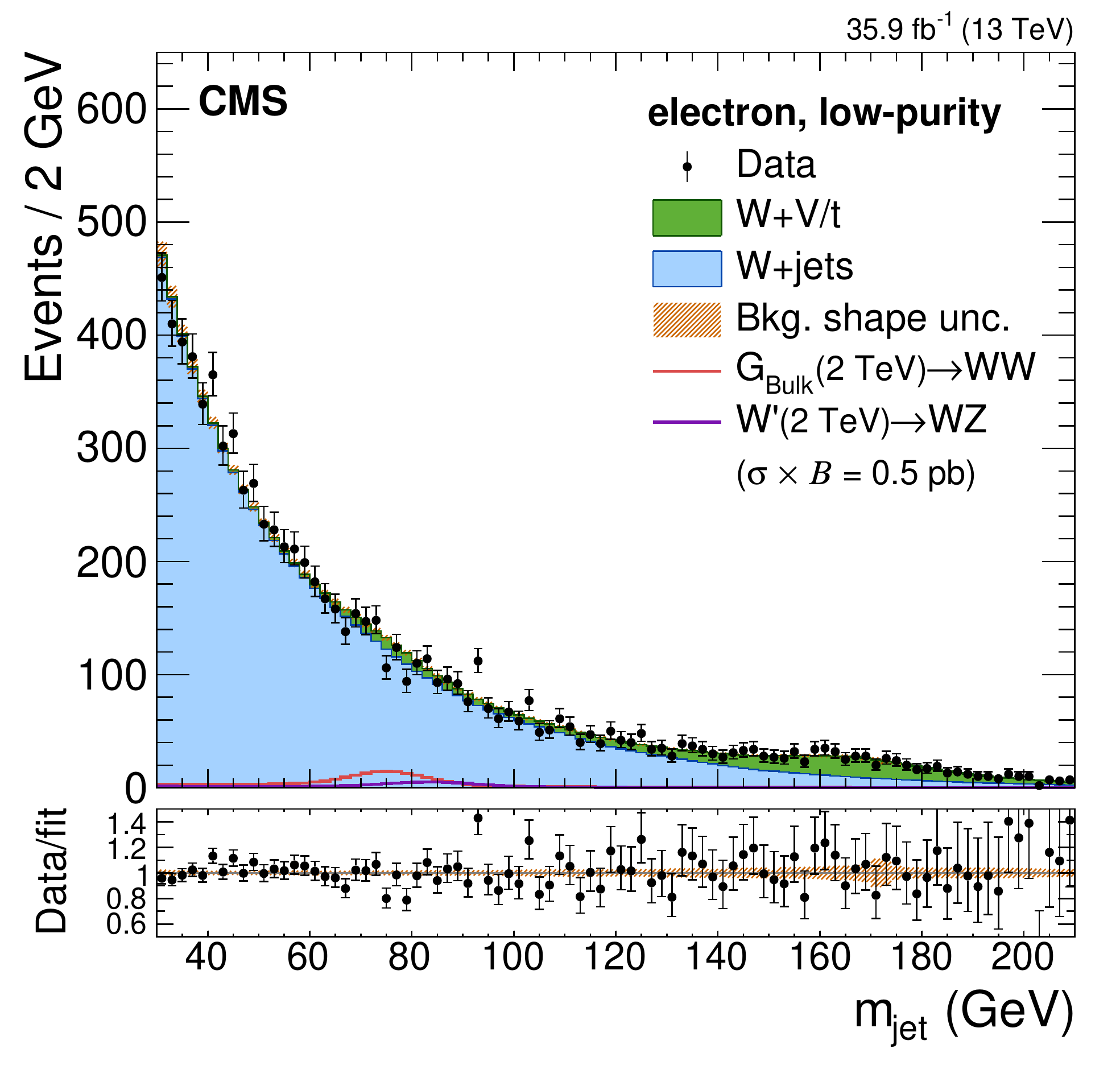}
		\includegraphics[width=0.45\textwidth]{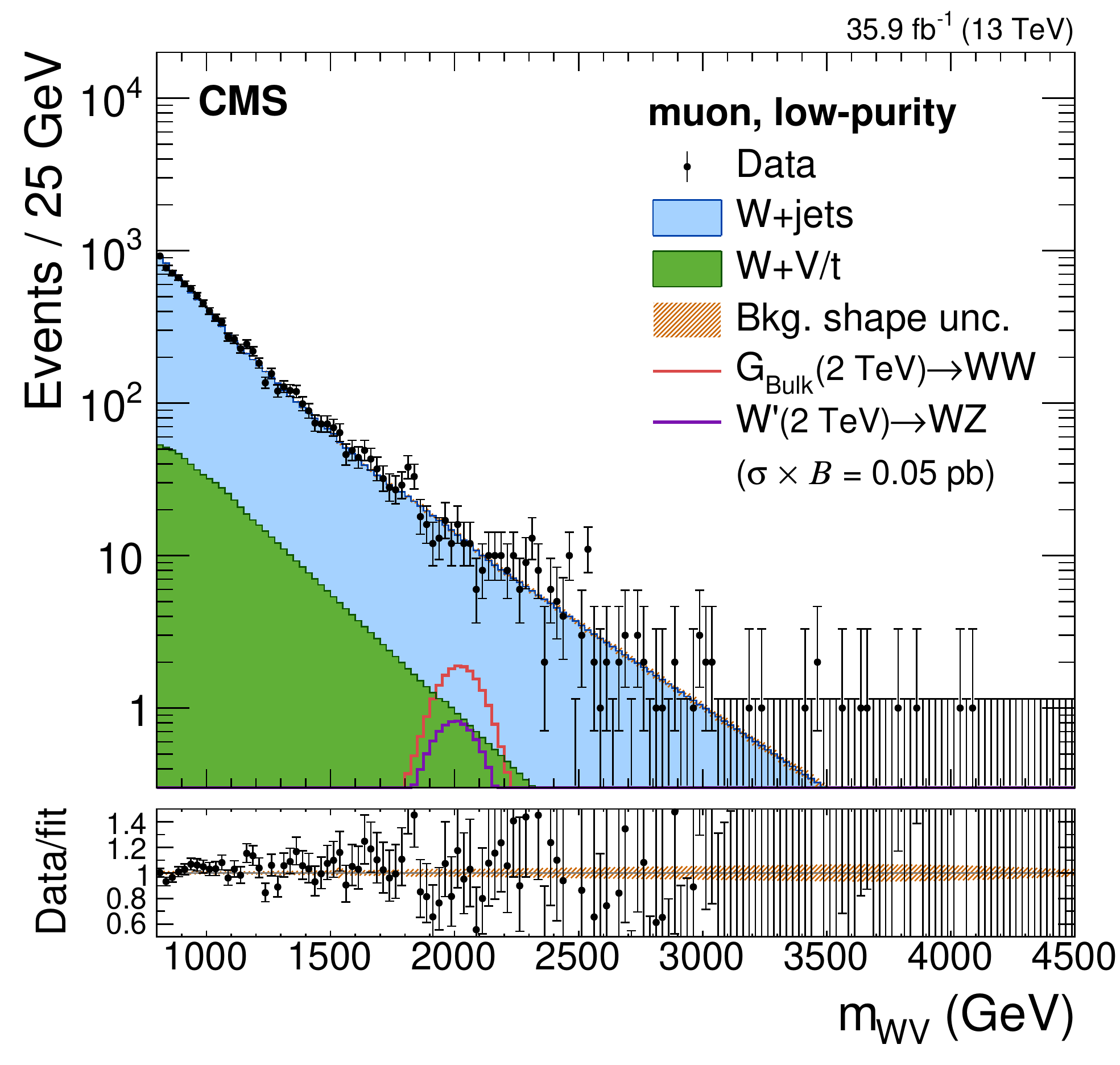}
		\includegraphics[width=0.45\textwidth]{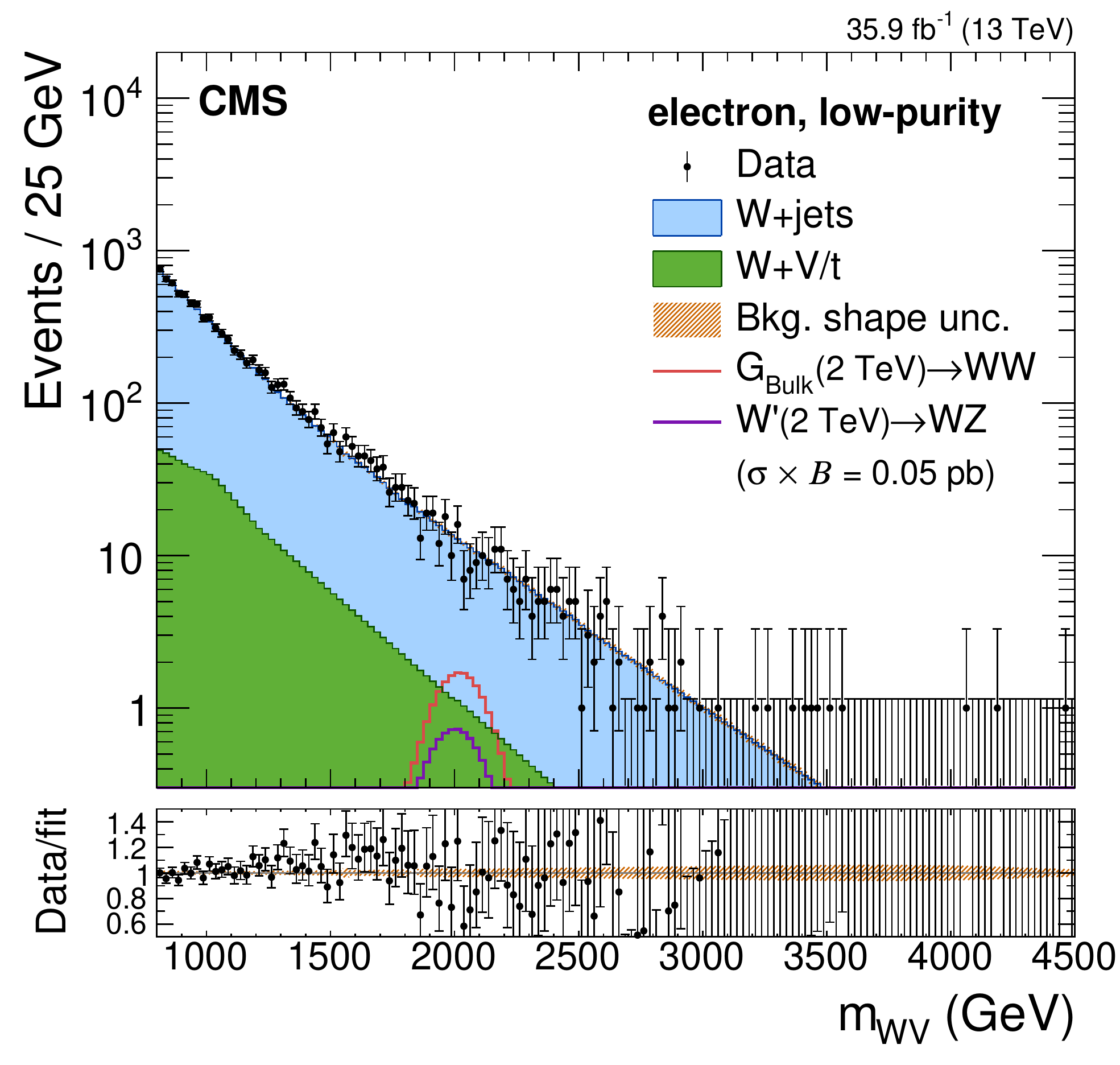}
	\caption{Comparison between the fit result and data distributions of $\mJ$ (upper) and $\mWV$ (lower)
		in the muon LP (left) and electron LP (right) category.
		The background shape uncertainty is shown as a shaded band, and the statistical uncertainties of the data are shown as vertical bars.
		No events are observed with $\mWV>4.5\TeV$.
		Example signal distributions are overlaid, using an arbitrary
		normalization that is different in the upper and lower plots.
		\label{fig:LP}}
\end{figure}

For both signal extraction techniques, the fit is performed simultaneously for the purity and
lepton flavour categories. The result of the two-dimensional fit is presented in Figs.~\ref{fig:HP} and \ref{fig:LP},
where projections in $\mJ$ and $\mWV$ are shown for the HP and LP categories, respectively.
The inclusive jet mass distributions demonstrate excellent modelling of both the resonant peaks and the continuum
for all categories. In the LP category, the resonant background is dominated by the merged top quark contribution.
These events show mostly a three-prong structure where both the quarks from the $\PW$ boson and the b quark are resolved
inside the large-radius jet, which makes it fail the $\nsubj$ HP requirement.

Good modelling of the data is also observed in the resonance mass projections, where the $\mWV$ distribution is plotted for events in the full jet mass range.
The best fit values of the nuisance parameters are found within the range initially associated with each uncertainty.
The quality of the fit is also quantified by forming a goodness-of-fit estimator using
the saturated model~\cite{Baker:1983tu}. The observed value of the estimator is compared
with toy experiments and falls within the 68\% interval containing the most probable results.

The search for new $\PW \PW$ and $\PW \PZ$ resonances is performed in the range of the resonance mass hypothesis between 1.0 and 4.4\TeV, which ensures that a hypothetical signal bump is contained within the $\mWV$ search region ranging from 0.8 to 5.0\TeV.
The data were also analyzed with the $\alpha$ method using separate $\PW$ and $\PZ$ boson mass windows.
The results are found to be consistent between the two methods within their respective uncertainties, taking into account correlations.

\begin{figure}[htbp]
	\centering
		\includegraphics[width=0.45\textwidth]{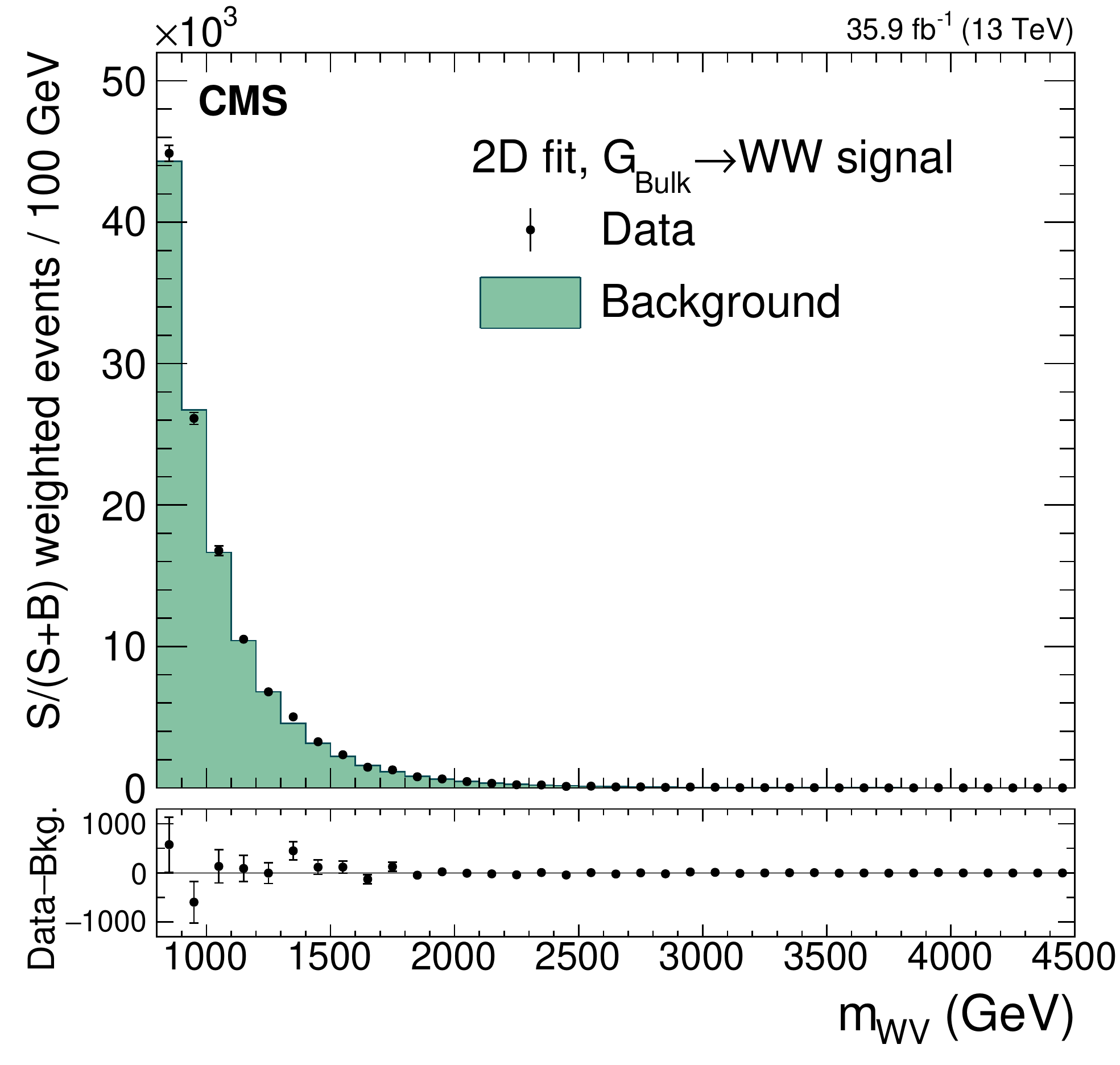}
		\includegraphics[width=0.45\textwidth]{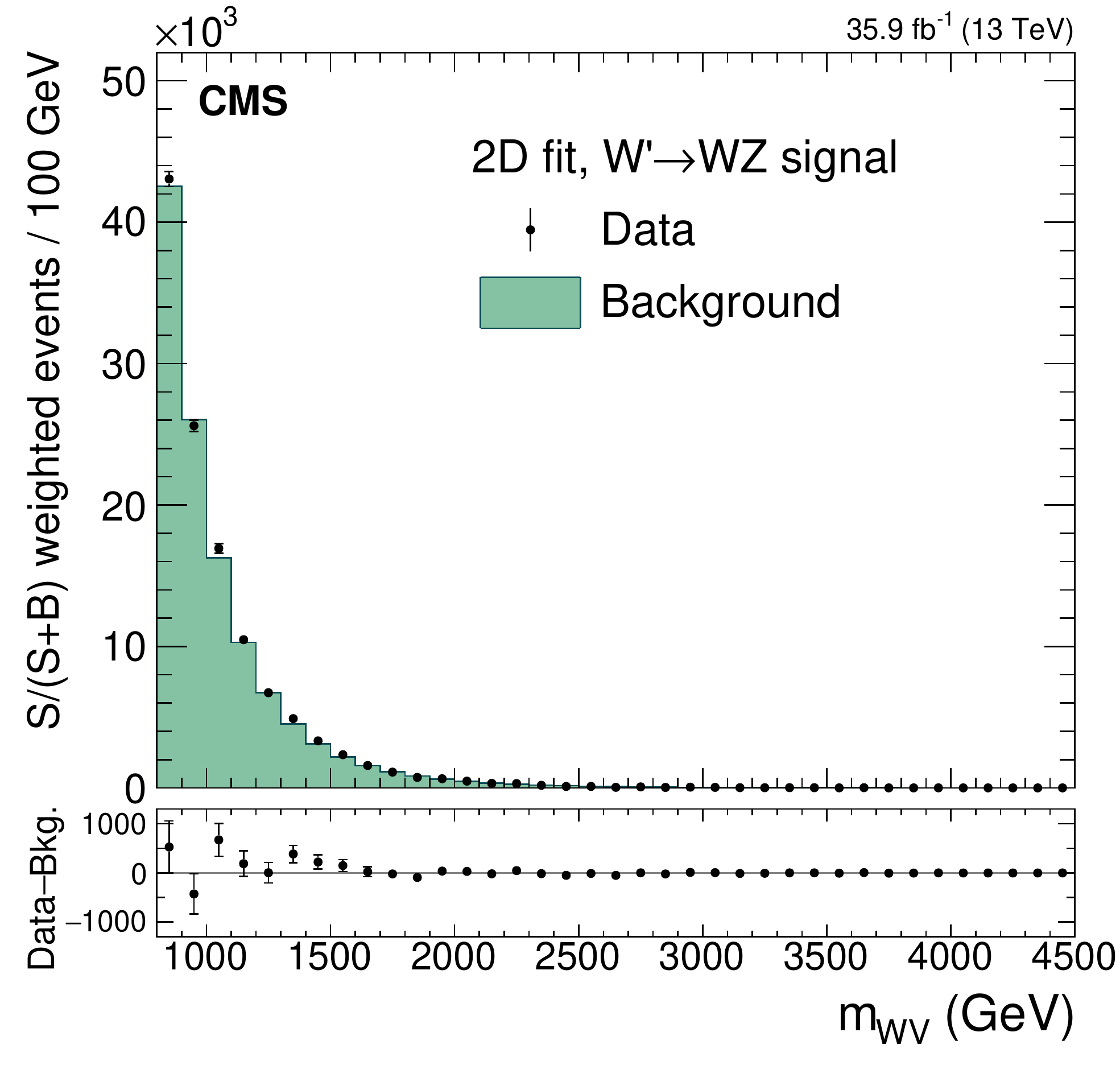}
		\includegraphics[width=0.45\textwidth]{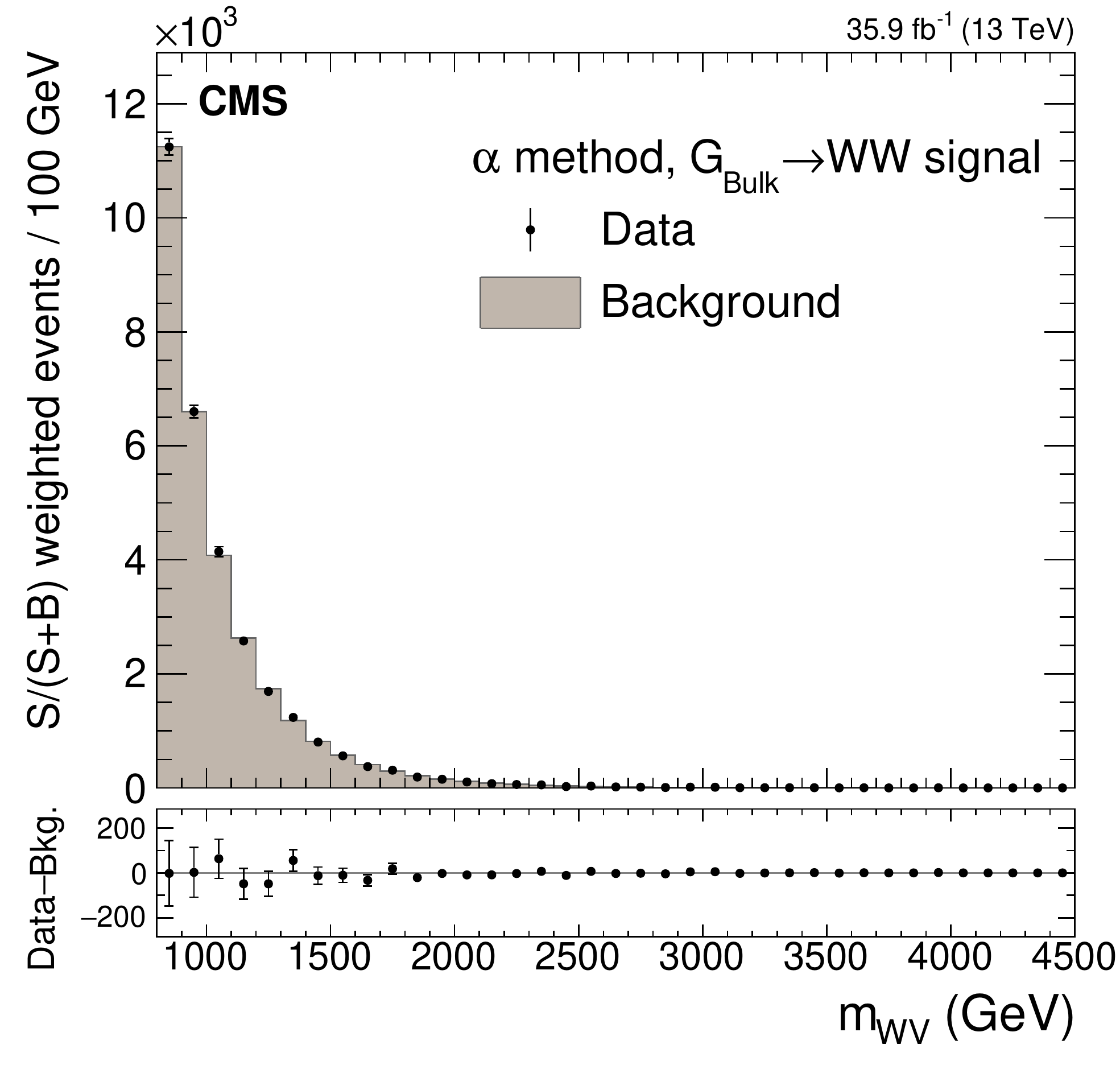}
		\includegraphics[width=0.45\textwidth]{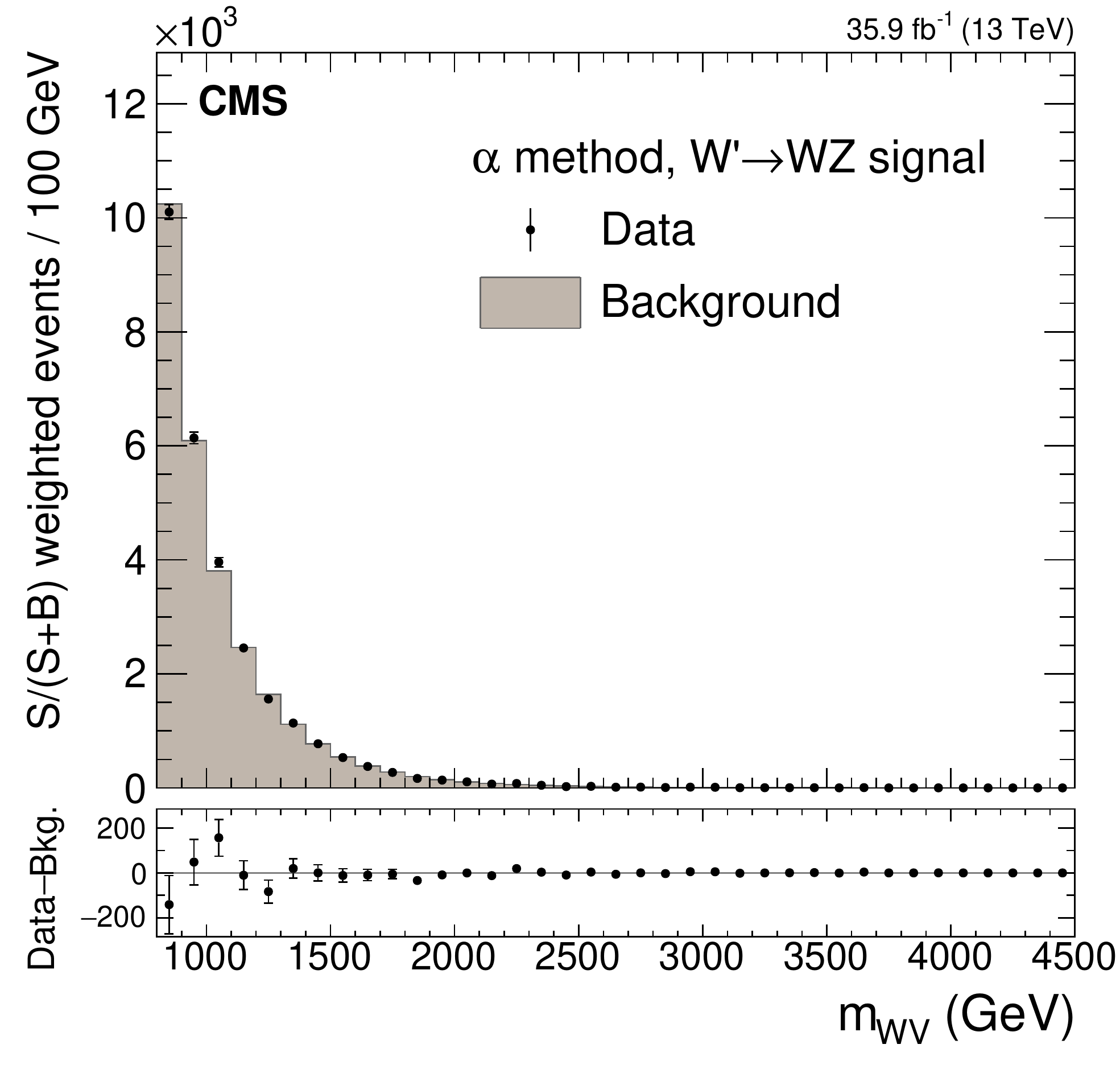}
	\caption{S/(S+B) event-weighted distributions of the resonance mass for the $\BulkG \to \PW\PW$ signal (left) and $\PWpr \to \PW \PZ$ signal (right) for the 2D fit (upper) and the $\alpha$ method (lower). The lower panels show the differences between the weighted data and the weighted fit results. The vertical bars correspond to the statistical uncertainties of the data.
		\label{fig:weighted}}
\end{figure}

Figure~\ref{fig:weighted} shows a summary of the contributions of all categories to the signal extraction. Each event
is weighted by the fraction of the number of expected signal events (S) over the sum of the expected signal and background events (S+B) in each category and in each
soft-drop jet mass bin. The signal is hereby normalized to the production cross section of a graviton or \PWpr\ of mass 2\TeV as predicted by the bulk graviton and HVT models, respectively, with parameters as defined in Section~\ref{sec:models}.

The largest deviation from the background hypothesis is observed for a mass hypothesis around 1.4\TeV with a
local significance of 2.4 (2.5) standard deviations for the $\PW\PW$ ($\PW\PZ$) signal,
while the global significance in the search range is found to be 1.2 (1.4) standard deviations.

Several additional checks were pursued to validate the new background estimation technique.
The range of the fit was reduced, starting at 1.0 instead of 0.8\TeV. The lower $\mWV$ region
is very important to constrain the top quark background around the $\PW$ jet mass peak.
After reducing the range, the observed and expected local significance was consistently lower, because of the loss in \ttbar{} event count.
Another test was to replace the 2D fit with a 1D binned fit on $\mWV$ after introducing soft-drop jet mass windows
similar to the $\alpha$ method. This test also yielded a background estimation compatible with the 2D fit result and a maximum local significance of 2.5
standard deviations.

\begin{figure}[h]
	\centering
		\includegraphics[width=0.45\textwidth]{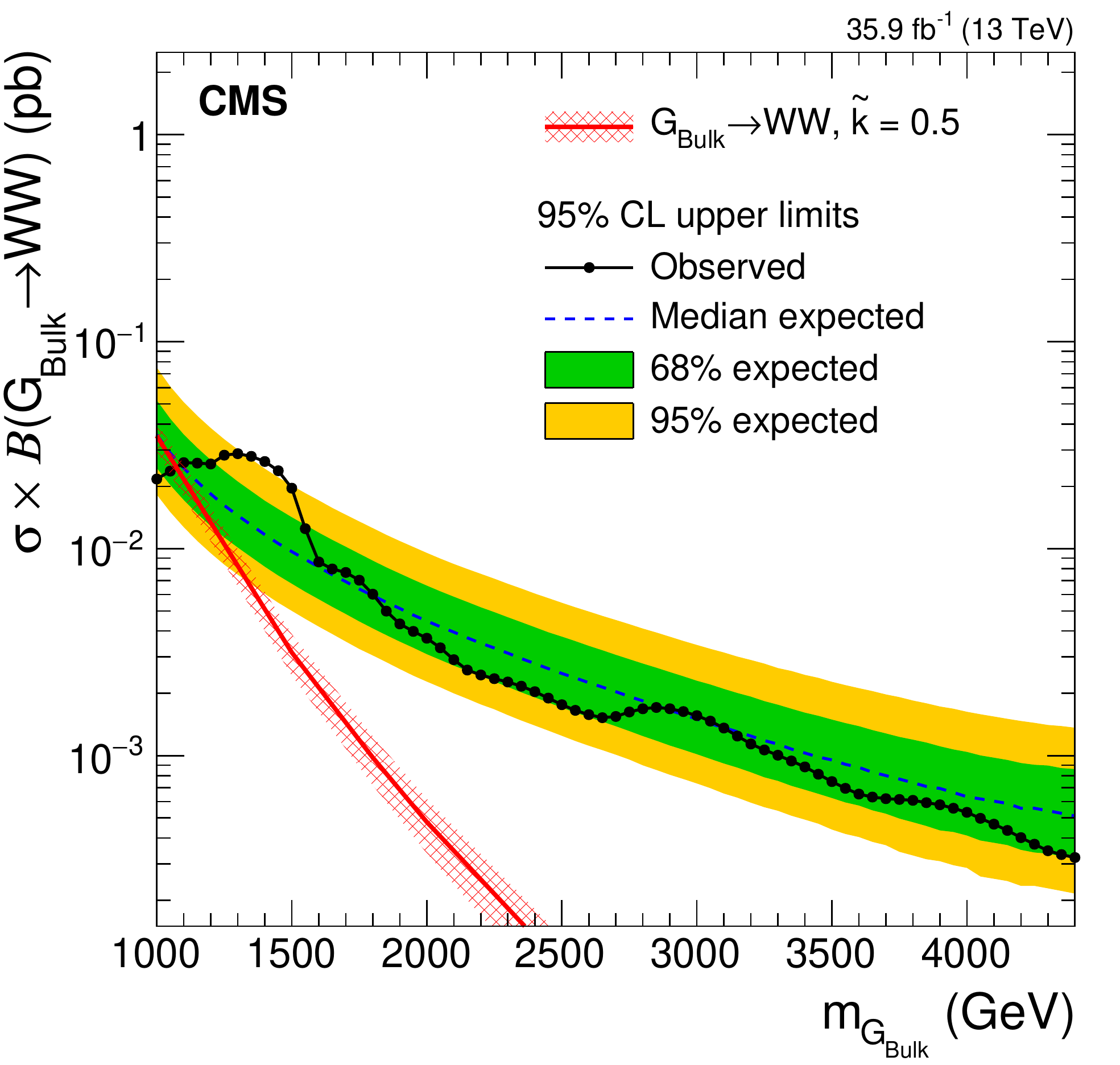}
		\includegraphics[width=0.45\textwidth]{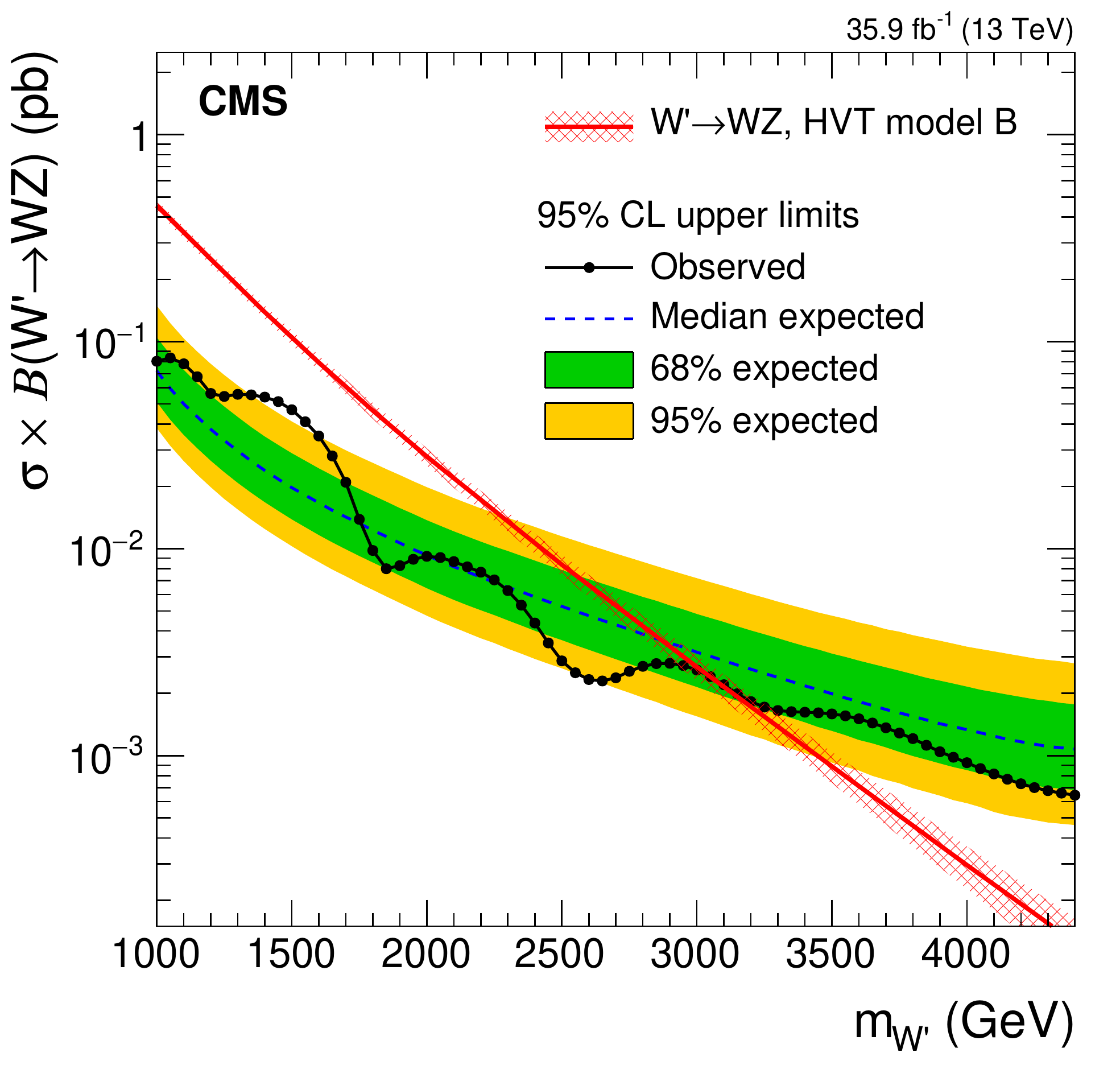}
	\caption{Exclusion limits on the product of the production cross section and the branching fraction for a new spin-2 resonance decaying to $\PW\PW$
		(left) and for a new spin-1 resonance decaying to $\PW\PZ$ (right), as a
		function of the resonance mass
		hypothesis. Signal cross section uncertainties are shown as red cross-hatched bands.\label{fig:exclusion_limits}}
\end{figure}

The results are interpreted in terms of exclusion limits for the
benchmark signal models described in Section~\ref{sec:models}.
We provide model-independent limits, which are not coupled to the relative normalizations of the benchmark models. 
We expect any model-dependent effects on the acceptance and selection efficiency to be covered by the PDF and scale uncertainties.
Figure~\ref{fig:exclusion_limits} shows the upper exclusion limits on the product of the resonance production cross section
and the branching fraction to $\PW \PW$ or $\PW \PZ$ as a function of the resonance mass.
The observed limits for the $\PW \PW$ signal range from 29\unit{fb} at 1.3\TeV to 0.32\unit{fb} at 4.4\TeV,
while for the $\PW\PZ$ signal they range from 84\unit{fb} at 1.05\TeV to 0.64\unit{fb} at 4.4\TeV.
By comparing these limits to the expected cross sections from the benchmark theoretical models,
$\PW \PW$ resonances lighter than 1.07\TeV and $\PW \PZ$ resonances lighter than 3.05\TeV  are excluded at 95\% confidence
level (\CL), using the asymptotic approximation~\cite{Cowan:2010js} of the
\CLs\ method~\cite{Junk:1999kv,Read:2002hq}.

\section{Summary}
\label{sec:end}
A search for a new heavy resonance decaying to a pair of vector bosons
is performed in events with one muon or electron and a massive
jet. Using the $N$-subjettiness ratio $\nsubj$, massive jets are
tagged as highly energetic vector bosons ($\PV=\PW,\PZ$) decaying to
quark pairs. The soft-drop mass is used as an estimate of the $\PV$-jet
mass. The lepton momentum and missing transverse momentum are used to reconstruct
the momentum of the $\PW \to \ell \Pgn$ boson candidate, constraining the
invariant mass of the $\ell \Pgn$ pair to the $\PW$ boson mass value. A novel
signal extraction technique is introduced based on a simultaneous fit of the
$\PV$-jet mass and the diboson mass, and improves the search sensitivity compared to the method employed in previous
versions of this analysis. No significant evidence of a new signal is found.
The results are  interpreted in terms of upper
limits on the production cross section of new resonances decaying to $\PW\PW$ and $\PW\PZ$ final states.
The observed limits for a $\PW \PW$ resonance range from 29\unit{fb} at 1.3\TeV to 0.32\unit{fb} at 4.4\TeV,
while for a $\PW \PZ$ resonance they range from 84\unit{fb} at 1.05\TeV to 0.64\unit{fb} at 4.4\TeV.
Comparing the excluded cross section values to the
expectations from theoretical calculations, spin-2 bulk graviton resonances decaying to $\PW \PW$ with mass smaller than 1.07\TeV and $\PWpr \to \PW \PZ$
resonances lighter than 3.05\TeV are excluded at 95\% \CL.

\begin{acknowledgments}
We congratulate our colleagues in the CERN accelerator departments for the excellent performance of the LHC and thank the technical and administrative staffs at CERN and at other CMS institutes for their contributions to the success of the CMS effort. In addition, we gratefully acknowledge the computing centres and personnel of the Worldwide LHC Computing Grid for delivering so effectively the computing infrastructure essential to our analyses. Finally, we acknowledge the enduring support for the construction and operation of the LHC and the CMS detector provided by the following funding agencies: BMWFW and FWF (Austria); FNRS and FWO (Belgium); CNPq, CAPES, FAPERJ, and FAPESP (Brazil); MES (Bulgaria); CERN; CAS, MoST, and NSFC (China); COLCIENCIAS (Colombia); MSES and CSF (Croatia); RPF (Cyprus); SENESCYT (Ecuador); MoER, ERC IUT, and ERDF (Estonia); Academy of Finland, MEC, and HIP (Finland); CEA and CNRS/IN2P3 (France); BMBF, DFG, and HGF (Germany); GSRT (Greece); OTKA and NIH (Hungary); DAE and DST (India); IPM (Iran); SFI (Ireland); INFN (Italy); MSIP and NRF (Republic of Korea); LAS (Lithuania); MOE and UM (Malaysia); BUAP, CINVESTAV, CONACYT, LNS, SEP, and UASLP-FAI (Mexico); MBIE (New Zealand); PAEC (Pakistan); MSHE and NSC (Poland); FCT (Portugal); JINR (Dubna); MON, RosAtom, RAS, RFBR and RAEP (Russia); MESTD (Serbia); SEIDI, CPAN, PCTI and FEDER (Spain); Swiss Funding Agencies (Switzerland); MST (Taipei); ThEPCenter, IPST, STAR, and NSTDA (Thailand); TUBITAK and TAEK (Turkey); NASU and SFFR (Ukraine); STFC (United Kingdom); DOE and NSF (USA).

\hyphenation{Rachada-pisek} Individuals have received support from the Marie-Curie programme and the European Research Council and Horizon 2020 Grant, contract No. 675440 (European Union); the Leventis Foundation; the A. P. Sloan Foundation; the Alexander von Humboldt Foundation; the Belgian Federal Science Policy Office; the Fonds pour la Formation \`a la Recherche dans l'Industrie et dans l'Agriculture (FRIA-Belgium); the Agentschap voor Innovatie door Wetenschap en Technologie (IWT-Belgium); the F.R.S.-FNRS and FWO (Belgium) under the ``Excellence of Science - EOS" - be.h project n. 30820817; the Ministry of Education, Youth and Sports (MEYS) of the Czech Republic; the Council of Science and Industrial Research, India; the HOMING PLUS programme of the Foundation for Polish Science, cofinanced from European Union, Regional Development Fund, the Mobility Plus programme of the Ministry of Science and Higher Education, the National Science Center (Poland), contracts Harmonia 2014/14/M/ST2/00428, Opus 2014/13/B/ST2/02543, 2014/15/B/ST2/03998, and 2015/19/B/ST2/02861, Sonata-bis 2012/07/E/ST2/01406; the National Priorities Research Program by Qatar National Research Fund; the Programa Estatal de Fomento de la Investigaci{\'o}n Cient{\'i}fica y T{\'e}cnica de Excelencia Mar\'{\i}a de Maeztu, grant MDM-2015-0509 and the Programa Severo Ochoa del Principado de Asturias; the Thalis and Aristeia programmes cofinanced by EU-ESF and the Greek NSRF; the Rachadapisek Sompot Fund for Postdoctoral Fellowship, Chulalongkorn University and the Chulalongkorn Academic into Its 2nd Century Project Advancement Project (Thailand); the Welch Foundation, contract C-1845; and the Weston Havens Foundation (USA).
\end{acknowledgments}

\bibliography{auto_generated}
\cleardoublepage \appendix\section{The CMS Collaboration \label{app:collab}}\begin{sloppypar}\hyphenpenalty=5000\widowpenalty=500\clubpenalty=5000\textbf{Yerevan Physics Institute,  Yerevan,  Armenia}\\*[0pt]
A.M.~Sirunyan, A.~Tumasyan
\vskip\cmsinstskip
\textbf{Institut f\"{u}r Hochenergiephysik,  Wien,  Austria}\\*[0pt]
W.~Adam, F.~Ambrogi, E.~Asilar, T.~Bergauer, J.~Brandstetter, E.~Brondolin, M.~Dragicevic, J.~Er\"{o}, A.~Escalante Del Valle, M.~Flechl, M.~Friedl, R.~Fr\"{u}hwirth\cmsAuthorMark{1}, V.M.~Ghete, J.~Grossmann, J.~Hrubec, M.~Jeitler\cmsAuthorMark{1}, A.~K\"{o}nig, N.~Krammer, I.~Kr\"{a}tschmer, D.~Liko, T.~Madlener, I.~Mikulec, E.~Pree, N.~Rad, H.~Rohringer, J.~Schieck\cmsAuthorMark{1}, R.~Sch\"{o}fbeck, M.~Spanring, D.~Spitzbart, A.~Taurok, W.~Waltenberger, J.~Wittmann, C.-E.~Wulz\cmsAuthorMark{1}, M.~Zarucki
% ...
\vskip\cmsinstskip
% ...
1:~~Also at Vienna University of Technology, Vienna, Austria\\
% ...

\end{sloppypar}
\end{document}